\documentclass[reprint,amsmath,twocolumn,amssymb, nobibnotes, aps, pra, superscriptaddress]{revtex4-1}

\usepackage{graphicx}
\usepackage{dcolumn}
\usepackage{bm}
\usepackage{color}

\usepackage{tikz}
\usetikzlibrary{shapes}
\usetikzlibrary{backgrounds}
\usetikzlibrary{plotmarks}
\usepackage{multirow}

\begin{document}
\def \beq{\begin{equation}}
\def \eeq{\end{equation}}
\def \bea{\begin{eqnarray}}
\def \eea{\end{eqnarray}}
\def \bem{\begin{displaymath}}
\def \eem{\end{displaymath}}
\def \P{\Psi}
\def \Pd{|\Psi(\boldsymbol{r})|}
\def \Pds{|\Psi^{\ast}(\boldsymbol{r})|}
\def \Po{\overline{\Psi}}
\def \bs{\boldsymbol}
\def \bl{\bar{\boldsymbol{l}}}
\newcommand{\ihat}{\hat{\textbf{\i}}}
\newcommand{\jhat}{\hat{\textbf{\j}}}

\title{{Discrete Approximation of  Topologically Protected Modes in Magneto-Optical Media}}
\author{Mark J. Ablowitz and Justin T. Cole}
\affiliation{Department of Applied Mathematics, University of Colorado, Boulder, Colorado 80309}
\date{\today}     
\begin{abstract}

 Topologically protected waves in {the} {linearly} polarized Maxwell's equations {with} { 
 gyrotropic/magneto-optic media} were studied a decade ago both computationally and experimentally. This paper develops a robust tight-binding model for this system that makes careful use of Wannier function representations. {The model provides very good approximations to the underlying band structure.
 When} solved on a semi-infinite strip, {it} produces exponentially localized edge modes whose corresponding eigenvalues span the frequency band gaps. A set of coupled differential equations are derived which {allows one to find how} the electromagnetic field propagates unidirectionally{, without} backscatter {from} defects. {Furthermore, the} discrete model predicts {topologically protected edge modes with} nontrivial Chern number which are consistent with direct simulation.

\end{abstract}

\pacs{Valid PACS appear here}

\maketitle

\section{Introduction}

Topological insulators in photonic lattices are an {exciting and active area of current research}. 
Among the distinguished features of these systems are the presence of the topologically protected edge modes that under the right set of conditions propagate unidirectionally{, without backscatter from defects.} 
With these 
properties in mind  there {have {been} numerous investigations to understand the nature, key properties and scope of these topological modes} 
{cf. \cite{top_ins_review}} and {to} understand how these states can {potentially} be harnessed in future applications \cite{top_laser_exp, MO_laser}. 

The idea of synthesizing a topological insulator within an electromagnetic system governed by Maxwell's equations was  first proposed in the seminal work of Haldane and Raghu \cite{haldane1,haldane2}.  In those works a transverse electric (TE) field in the presence of a anisotropic permittivity tensor with periodic structure was considered. By introducing a set of {hermitian,} purely imaginary, off-diagonal elements into the permittivity tensor it is possible to break time-reversal symmetry. Doing this has the effect of creating a gap at the inter-band touching points,  which is filled by {so-called `gapless'  topologically protected} edge modes.
 {The underlying topological structure is} established by the existence of nontrivial topological invariants {such as} Chern numbers. {The spectrum and corresponding solutions of this problem were rigorously studied in \cite{leethorp}.} 

A different system, inspired by {the above work of Haldane and Raghu}, found transverse magnetic (TM) fields {that} {can} exhibit gapless edge states when an external magnetic field is applied to an array of ferrimagnet rods \cite{wang1}. The external bias generates a {magneto-optical (MO)} response that manifests itself in a gyrotropic permeability tensor \cite{pozar} which breaks time-reversal symmetry, but not inversion symmetry. As a result, the system 
 {possesses} nontrivial Chern invariants and 
support{s} unidirectional mode propagation analogous to the quantum Hall effect \cite{vonklitzing,TKNN}. 
 {T}hese modes {were} experimentally observed within a square lattice \cite{wang2}. A similar topological insulator system that uses {a} honeycomb lattice instead  was also found to support topologically protected modes \cite{ao,poo}.

{Subsequently a photonic} {topological insulator} {was investigated in} \cite{Rechtsman} {where} an effective magnetic field was induced by the helical driving of a honeycomb waveguide in the direction of propagation. Advantages of this system are that {no external magnetic field}  is required and the frequencies are in the optical regime. {A tight binding model of   this  system including nonlinear effects and propagating solitons along an edge was discussed in \cite{AblCCYM2014}}.  This system has been shown to support novel topological phenomena such as Weyl points \cite{Noh} and valley-hall states \cite{Noh2}.

The purpose of this paper is to construct a tight-binding model that describes the {MO system considered in} 
 \cite{wang1}
 {as well as  {a broad class of related problems.}}
To our knowledge no such tight-binding system {describing these type of MO systems have been found before. Tight-binding approximations produce {effective, simplified} discrete models;  they are able to capture the essence of the system, provide important qualitative information, and relative to full numerics they are computable at a fraction of the cost.
In this context we employ a novel sequence of methods that are appropriate for this application.}

Here we  employ a Wannier mode expansion of the electromagnetic field to derive our topological insulator model; {this is different from} an expansion in terms of gaussian-type orbitals. The latter, while very useful for Schr\"odinger operators cf. \cite{AC1,AC2}, have not been found to be effective in this topological class of problems{.  
The difficulty} with using a direct Wannier mode expansion is that the {MO} system is a Chern insulator and as a result the decay rate of the Wannier functions {is  slow  \cite{brouder}. This means that  a direct Wannier approach  to obtain a tight-binding system for a Chern insulator  (i.e. a system with nontrivial Chern numbers which occurs in this problem due to the breaking of time-reversal symmetry)} {presents serious obstacles.}

To obtain a localized basis of modes from which to construct our discrete model we employ a {\it perturbative} Wannier approach whereby we obtain a set of Wannier modes from a closely related equation that {\it does} possess time-reversal symmetry. We find that this approximate set of Wannier modes produces spectral band diagrams which are in {remarkably} good agreement with numerics. From the above Wannier expansion we construct a system of differential equations {for the {modal} coefficients} from which we compute the spectral bands and corresponding edge eigenmodes. The adjacent bulk bands are found to possesses nontrivial Chern numbers which, by the bulk-edge correspondence, indicate the presence of topologically protected modes \cite{hatsugai}. Moreover, evolution of these edge states is found to propagate unidirectionally around a lattice defect. {These results are in agreement with prior MO work \cite{wang1,wang2, lu_review}.}

The {outline of the paper is the following.} 
In Sec.~\ref{tm_maxwell_eqn_sec} {starting from Maxwell's equations} we formulate the TM equation that governs this MO system. 
 {We} approximate the ferrite rods by super-Gaussian functions and 
 {represent} the electric field by a Bloch wave. In Sec.~\ref{wannier_sec} we give a brief introduction to Wannier functions and motivate our perturbative {approach.}
With {this} Wannier {Galerkin-type} basis in hand, the {tight-binding} discrete system is derived{; the} coefficients {of the discrete model} {are} {calculated} in Sec.~\ref{TbA_sec} {and 
 the discrete system {is solved}} in Sec.~\ref{spec_band_sec}. There we compute the linear spectral bands for both the bulk and edge problems. The bulk bands are found to possess nonzero Chern numbers that {with the bulk-band correspondence} agree with the {number {and orientation} of} gapless edge modes we find. Finally, in Sec.~\ref{time_evolve_sec} direct numerical simulations are performed on the discrete model. The edge modes are found to flow unidirectionally and propagate {without backscatter} around lattice defects. We conclude in Sec.~\ref{conclude}.
 
\section{The TM Equation with Gyrotropic Tensor}
\label{tm_maxwell_eqn_sec}

We start at the source-free and current-free Maxwell's equations 
\begin{align}
\label{maxwell}
& \nabla \times {\bf H} = \frac{\partial {\bf D}}{\partial t}  , ~\nabla \times {\bf E} = - \frac{\partial {\bf B}}{\partial t}, ~ \nabla \cdot {\bf D} = 0, ~\nabla \cdot {\bf B} = 0 ,
\end{align}
where ${\bf H}$ is the magnetic field, ${\bf E}$ is the electric field, ${\bf D}$ is the electromagnetic displacement, and ${\bf B}$ is the magnetic induction.
Consider a real time-harmonic electromagnetic field ${\bf E}({\bf r} , t) = {\bf \widetilde{E}}({\bf r},\omega ) e^{i \omega t} + c.c. , ~{\bf H}({\bf r} , t) = {\bf \widetilde{H}}({\bf r} ,\omega) e^{i \omega t} + c.c. $, where $c.c.$ is the complex conjugate and $\omega$ is the angular frequency.

 {The} electric field {is taken to be} linearly polarized in the perpendicular $z$-direction and depends on the transverse variables $x,y$, i.e. ${\bf \widetilde{E}}({\bf r} ,\omega) = (0 ,0, E({\bf r}, \omega)) $ such that ${\bf r} = (x,y)$.
The displacement and electric fields are related via
\begin{equation}
\label{elec_relate}
{\bf {D}}({\bf r},t)  = {\epsilon}({\bf r}) {\bf{E}}({\bf r}, t) \; ,
\end{equation}
 {with} scalar permittivity function $ {\epsilon} ({\bf r}).$ {For a linearly polarized field this displacement vector clearly satisfies the divergence-free condition $\nabla \cdot {\bf D} = 0$.} The function ${\epsilon} ({\bf r})$ models a square array of YIG rods ($\epsilon = 15 \epsilon_0$) {such as} 
 those considered in \cite{wang1,wang2} [see Fig.~\ref{sq_lattice_plot}(b)]. 
 {A constant} external magnetic field $H_0 $ is applied in the $z$-direction.
For a ferrimagnetic material the external magnetic field aligns the magnetic dipoles and induces {a} 
magnetization response \cite{pozar}. The magnetic induction and magnetic field are 
 {related} by {the gyrotropic permeability tensor}
\begin{equation}
\label{relate_BH}
{\bf \widetilde{B}} = [\mu] {\bf \widetilde{H}} \; , ~~~~~~~~~~~
 [\mu] ({\bf r})= 
 \begin{pmatrix}
 \mu & i \kappa & 0 \\
 - i \kappa & \mu & 0 \\
 0 & 0 & \mu_0 
 \end{pmatrix}({\bf r}) \; ,
\end{equation}
where 
\begin{align*}
\label{magnetic_coeffs}
\mu= \mu_0 \left( 1 +  \frac{\omega_0 \omega_m }{\omega_0^2 - \omega^2 } \right) ~~~~~~~ {\rm and} ~~~~~~~ \kappa = \mu_0  \frac{\omega \omega_m }{\omega_0^2 - \omega^2 } \; .
\end{align*}
The frequencies above are {$\omega_0({\bf r}) = \mu_0 \gamma H_0({\bf r})$ and $\omega_m({\bf r})= \mu_0 \gamma M_s({\bf r})$, where $\mu_0$ is the vacuum permeability,}$\gamma = 1.759 \times 10^{7} \text{rad/(Gs)}$ is the gyromagnetic ratio and $M_s$ is the magnetization saturization. {We point out that the relationship between the magnetic fields in (\ref{relate_BH}) satisfies the divergence relation {$\nabla \cdot {\bf B} = 0$. }} 

For simplicity throughout this paper we assume a dispersionless tensor and 
fix the frequency $\omega$ in the coefficients $\mu$ and $\kappa${.} The dispersive problem was {considered}  in \cite{wang1}; it  did not significantly alter the structure of the spectral bands and resulting {topology}. 
We {use}  
typical physical parameters $H_0 = 1600 ~\text{Oe}$ and $ M_s = 1810~ \text{G}$ resulting in values of $\mu = 14 \mu_0$ and $\kappa = 12.4 \mu_0$ within the ferrite rods.
Combining equations (\ref{maxwell}), (\ref{elec_relate}), and (\ref{relate_BH}) yields
\begin{equation}
\label{inverse_tensor}
\nabla \times ([ \mu]^{-1} \nabla \times {\bf \widetilde{E}}  )  = \omega^2  {\epsilon}{\bf \widetilde{E}} \; ,
\end{equation}
where
\begin{equation*}
[\mu]^{-1} =
\begin{pmatrix}
\tilde{\mu}^{-1} & i \eta & 0 \\
- i \eta & \tilde{\mu}^{-1} & 0 \\
0 & 0 & \mu_0^{-1}
\end{pmatrix} \; ,
\end{equation*}
such that $\tilde{\mu} =  (\mu^2 - \kappa^2)/  \mu$ and $\eta = - \kappa / (\mu^2 - \kappa^2).$
For a linearly polarized field we obtain the following ``master'' equation
\begin{equation}
\label{master_eqn}
 - \nabla^2 E + [\nabla \ln \tilde{\mu} - i \tilde{\mu} ( \bm{\hat{z}} \times \nabla \eta ) ] \cdot \nabla E =   \omega^2 {\epsilon}  \tilde{\mu}  E \; . 
\end{equation}
This equation is non-dimensionalized by 
\begin{equation*}
{\bf r} = a {\bf r}'  , ~~~ \mu = \mu_0 \mu'  , ~~~ \kappa = \mu_0 \kappa' , ~~~ \epsilon = \epsilon_0 \epsilon' , ~~~ \omega = \frac{ c}{a} \omega'  ,
\end{equation*}
where $c = 1/\sqrt{\mu_0 \epsilon_0}$ is the speed of light in vacuum and $a$ is the distance between {two adjacent} lattice rods. The non-dimensionalized master equation (after dropping the $'$ notation) is 
\begin{align}
\label{master_eqn_non_dim}
 - & \nabla^2 E +  \mathcal{M}({\bf r}) \cdot \nabla E =   \omega^2 {\epsilon}  \tilde{\mu}  E ~ , \\ \nonumber
 & \mathcal{M}({\bf r}) \equiv \nabla \ln \tilde{\mu} - i \tilde{\mu} ( \bm{\hat{z}} \times \nabla \eta )  ~ .
\end{align}
{In the frequency domain, t}he imaginary term in Eq.~(\ref{master_eqn_non_dim}) is responsible for breaking time-reversal symmetry.  As we will see below, this term {leads to} 
bands with nontrivial Chern numbers and gapless edge modes.

\subsection{Periodic Lattice}
\label{lattice_approx_sec}

The  ferrite rods in the non-dimensionalized lattice {are taken to} have period $1$ in the transverse plane. 
The corresponding lattice vectors are
\begin{equation}
 {\bf e}_1 = (1,0) ~, ~~~~~~ {\bf e}_2 = (0,1) \; .
\end{equation}
Starting from the origin, the location of the rods are integer multiples of the lattice vectors, given by  {${\bf R}_{mn} \equiv  m {\bf e}_1 +  n  {\bf e}_2 $}, where $m,n \in \mathbb{Z}.$ The locations given by ${\bf R}_{mn}$ correspond to the centers of the lattice rods below; which we later refer to as the integer lattice sites. Both the permittivity and permeability tensors posses this periodic structure, hence they exhibit the following translation invariance
\begin{align*}
\epsilon({\bf r} + {\bf R}_{mn}) & = \epsilon({\bf r}) \; , ~~ \mu({\bf r} + {\bf R}_{mn}) = \mu({\bf r})  \; ,\\
 ~~&  \kappa({\bf r} + {\bf R}_{mn}) = \kappa({\bf r}) ~ .
\end{align*}
To model a single YIG rod of the lattice we {use the following}  super-Gaussian function 
\begin{equation}
\label{super_gauss}
l(x,y) = \exp\left[- \left( \frac{\sqrt{x^2 + y^2}}{\delta} \right)^6 \right] \; .
\end{equation}
This function {approximates} a cylindrical rod of radius $0.11$ considered in \cite{wang1}. To make the two functions comparable we choose the parameter $\delta$ so that the 
{half width at half maximum}
of (\ref{super_gauss}) is equal to the radius of the cylindrical rods; {this yields} $\delta = 0.11/(\log 2)^{1/6}$. To obtain band diagrams that {are close} those found using a pure cylinder, it is important this function {has a sharp derivative}.

\begin{figure} [h]
\centering
\includegraphics[scale=0.35]{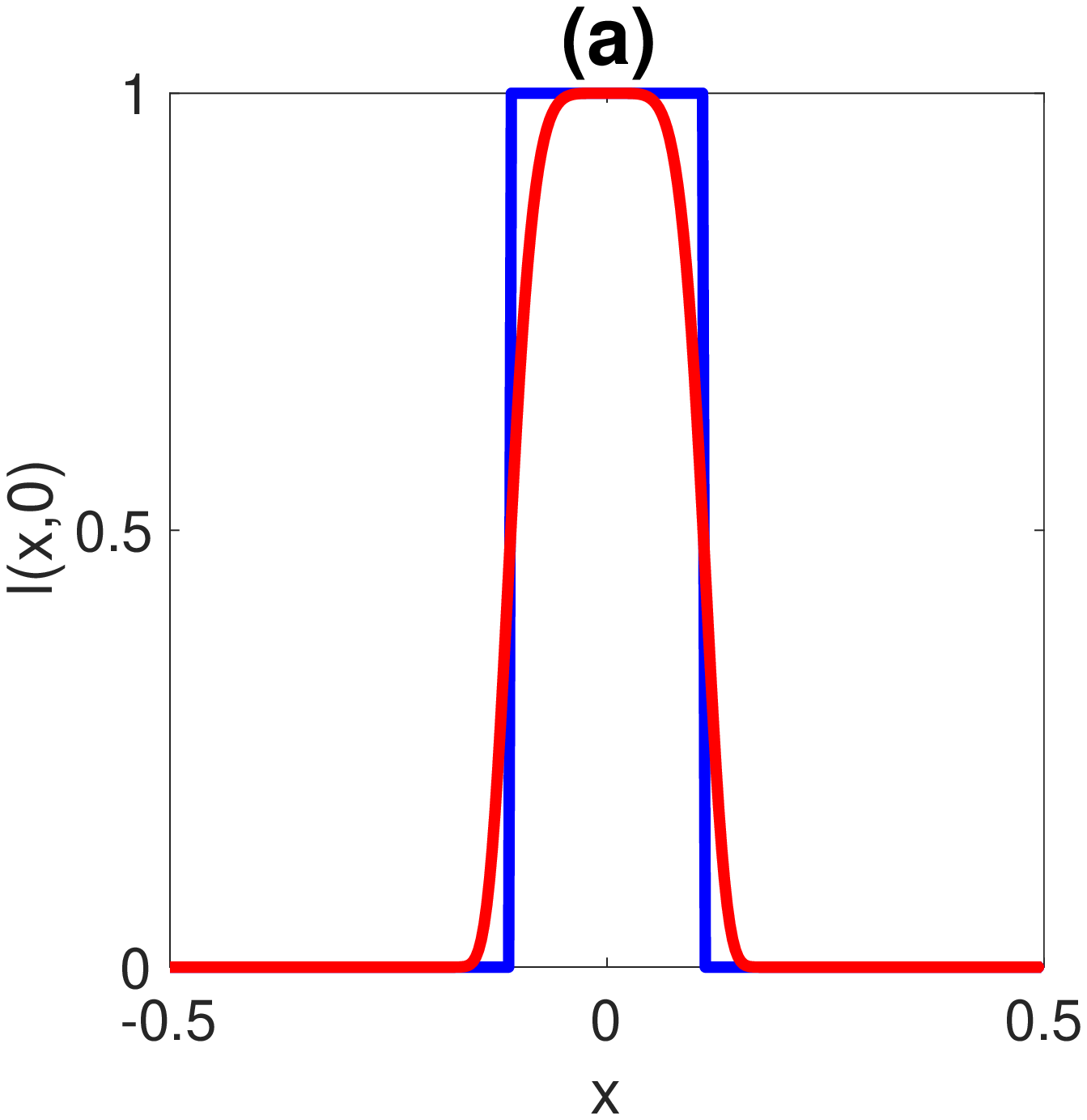}
\includegraphics[scale=0.35]{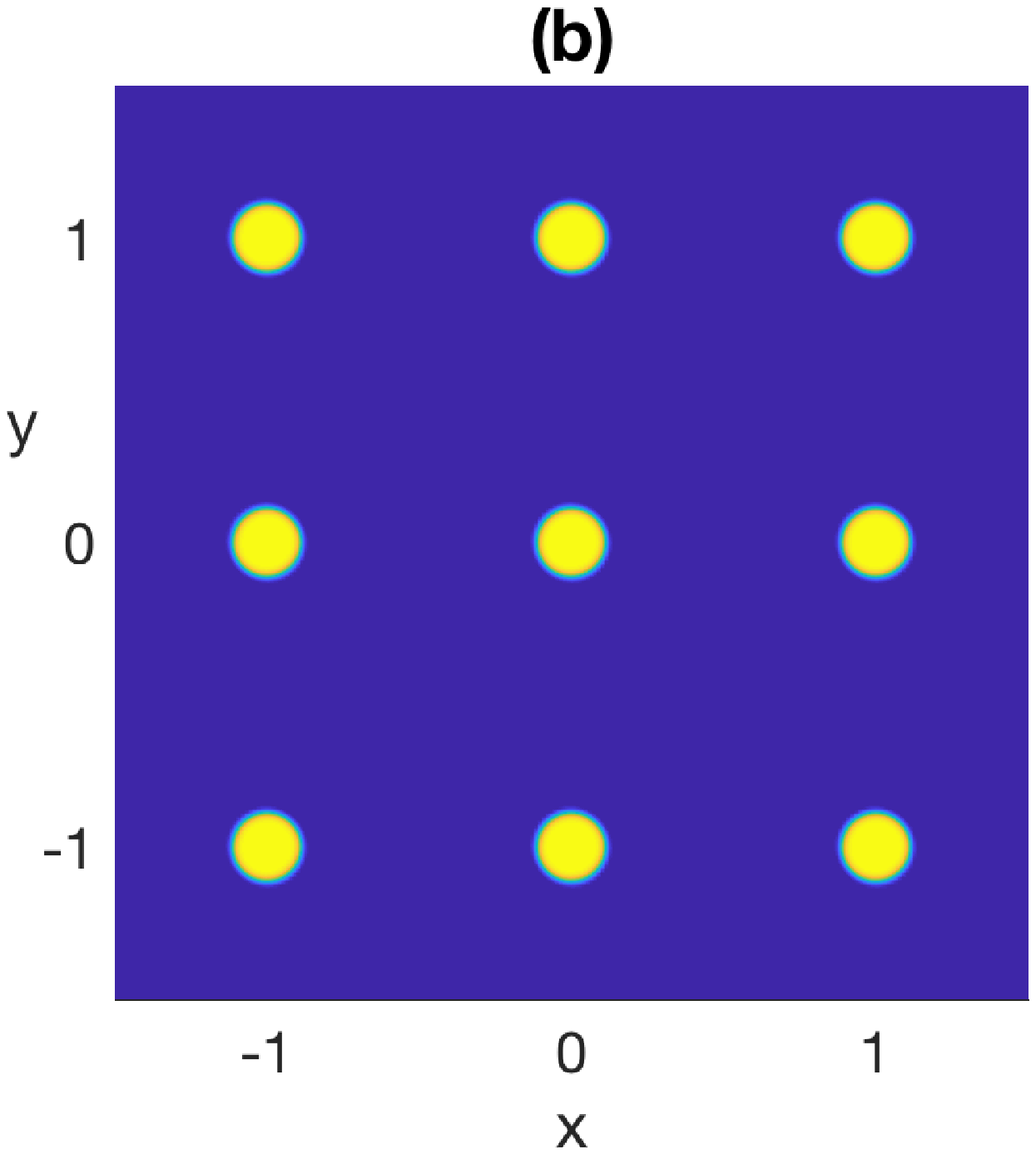}
\includegraphics[scale=0.35]{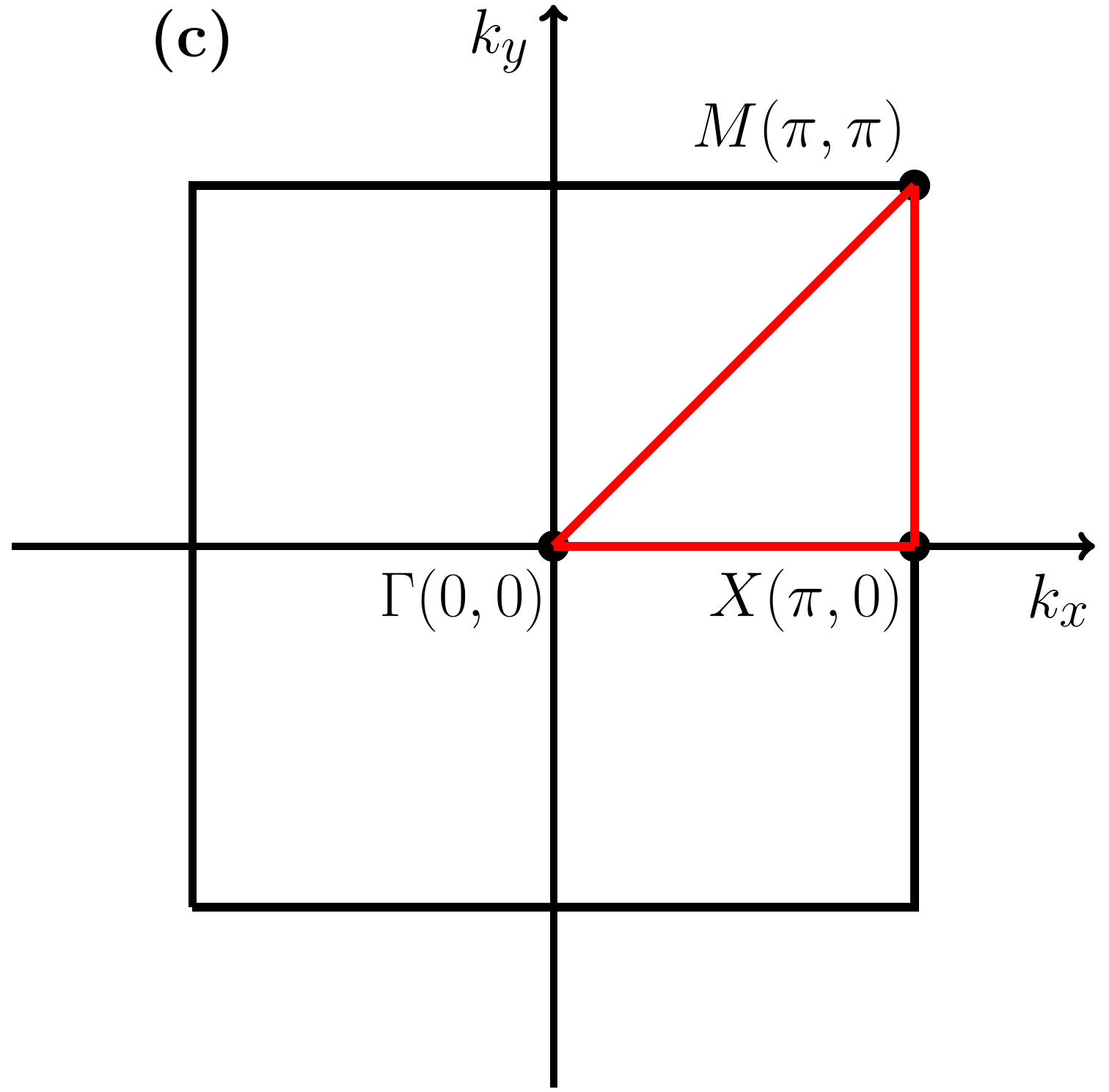}
\caption{(color online) (a) Super-Gaussian profile (red curve) given in (\ref{super_gauss}) in comparison to cylinder of radius $0.11$ profile (blue curve). (b) Square array of super-Gaussian rods. (c) First Brillouin zone (BZ) of the square lattice: $[-\pi,  \pi] \times [-\pi ,  \pi]$. The irreducible Brillouin zone is the triangular region with vertices $\Gamma, X, M$. }
\label{sq_lattice_plot}
\end{figure}

To model an array of rods we form a series of super-Gaussians centered at the integer lattices sites.  The explicit functions we use are
\begin{align}
\nonumber
\epsilon({\bf r}) = 1 + 14 \sum_{m,n} l(x-m,y-n) \; , \\ \label{potential_model}
 \mu({\bf r}) = 1 + 13 \sum_{m,n} l(x-m,y-n) \; ,  \\ \nonumber
  \kappa({\bf r}) = 12.4 \sum_{m,n} l(x-m,y-n) \; .
\end{align}
A profile comparison between a cylindrical rod and the super Gaussian in (\ref{super_gauss}) is shown in Fig.~\ref{sq_lattice_plot}(a). A plot of the arrays in (\ref{potential_model}) is shown in Fig.~\ref{sq_lattice_plot}(b). {These periodic arrays are used to compute the Bloch modes discussed in the next section. In turn, the Bloch modes are used to directly compute bands and  the Wannier functions.}

\subsection{Numerical Computation of Bloch Eigenfunctions}
\label{num_comp_spec_bands}

%
%

Consider the sourceless eigenvalue problem in (\ref{master_eqn_non_dim}) with periodic coefficients.
Solutions are Bloch modes of the form {
\begin{equation}
\label{bloch_wave_define}
E_p({\bf r}, {\bf k}) = e^{i {\bf k} \cdot {\bf r}} u_p({\bf r}, {\bf k}) ~ , ~~~~ u_p({\bf r} + {\bf R}_{mn}, {\bf k}) = u_p({\bf r}, {\bf k}) ~ ,
\end{equation}
where $p$ denotes the corresponding band number and} ${\bf k} \equiv (k_x,k_y)$ is the 2D quasimomentum. This then gives {
\begin{equation}
\label{master_eqn_bloch_sub}
- (\nabla + i {\bf k})^2 u_p  + \mathcal{M}({\bf r}) \cdot (\nabla + i {\bf k}) u_p = \omega^2({\bf k})  \epsilon({\bf r}) \tilde{\mu}({\bf r}) u_p  ~ .
\end{equation}
} Since the function {$u_p({\bf r}, {\bf k}) $} is periodic in ${\bf r}$ it {can be} expanded in the Fourier series 
\begin{equation}
\label{U_Fourier}
u_p({\bf r}, {\bf k}) = \sum_{m,n} \widehat{u}_{p,mn}({\bf k}) e^{i {\bf G}_{mn} \cdot {\bf r}} ~ ,
\end{equation}
where ${\bf G}_{mn} \equiv 2\pi m {\bf e}_1 + 2\pi n {\bf e}_2  , ~ m,n \in \mathbb{Z} .$ The effective permittivity and permeability functions are also periodic and expanded in terms of Fourier series:
\begin{align*}
(\epsilon \tilde{\mu}) ({\bf r}) &= \sum_{m,n} \widehat{(\epsilon \tilde{\mu})}_{{mn}} e^{i {\bf G}_{mn} \cdot {\bf r}} \; , \\ 
\mathcal{M} ({\bf r}) & = \sum_{m,n} \widehat{\mathcal{M}}_{{mn}} e^{i {\bf G}_{mn} \cdot {\bf r}} .
\end{align*}
 {For this spectral method to be effective} 
 {the} potential in (\ref{potential_model}) {needs to be} smooth enough that the Fourier coefficients decay sufficiently rapidly. On the other hand, the magnetic term $\mathcal{M}({\bf r})$ in (\ref{master_eqn_non_dim}) {needs} to be strong enough ({i.e. have a large enough} gradient) to open a sufficiently large gap. We find that the super-Gaussian in (\ref{super_gauss}) is a compromise between these opposing sides. For the Bloch modes computed in this paper we took $40$ Fourier modes and a spatial discretization of $\Delta x = \Delta y = 1/240.$

Eigenvalue problem (\ref{master_eqn_bloch_sub}) is transformed into the coupled, algebraic system of equations 
\begin{align}
\label{Eig_Fourier_Form}
&  ({\bf G}_{mn} +  {\bf k})^2 \widehat{u}_{p,mn} ({\bf k}) \\ \nonumber 
+& i \sum_{m',n'} \widehat{\mathcal{M}}_{m'n'} \cdot ({\bf G}_{m-m',n-n'} +  {\bf k}) \widehat{u}_{p,m-m',n-n'}({\bf k}) \\ \nonumber
 &= \omega^2({\bf k})  \sum_{m',n'} \widehat{(\epsilon \tilde{\mu})}_{m'n'}({\bf k}) \widehat{u}_{p,m-m',n-n'}({\bf k})  \; ,
\end{align}
\begin{figure} [h]
\centering
\includegraphics[scale=0.31]{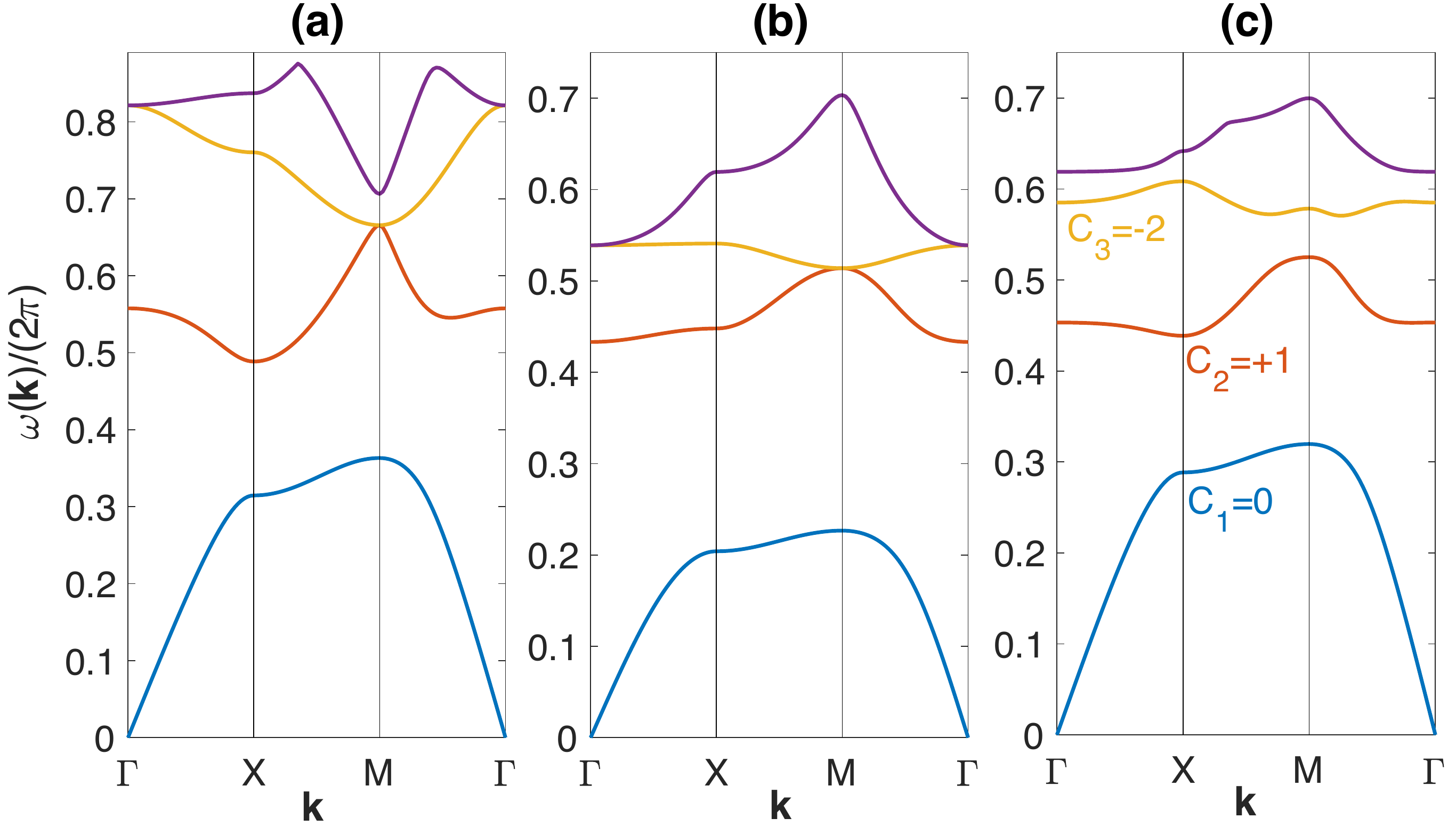}
\caption{First four dispersion bands in ascending order. Values of ${\bf k}$ are taken along the perimeter of the irreducible zone in Fig.~\ref{sq_lattice_plot}. Bands {are computed from} (\ref{master_eqn_bloch_sub}) {using parameters:} (a) $\tilde{\mu} = 1, \mathcal{M}({\bf r}) = {\bf 0}$, (b) $\tilde{\mu}$ defined in {(\ref{inverse_tensor}) and (\ref{potential_model})}, $\mathcal{M}({\bf r}) = {\bf 0}$, and (c) $\tilde{\mu}$ defined in {(\ref{inverse_tensor}) and (\ref{potential_model})}, $\mathcal{M}({\bf r}) \not= {\bf 0}$. {The computed Chern numbers {defined   in (\ref{chern_define}) for the corresponding bands}  are included in (c).}}
\label{spectral_curves}
\end{figure}
where $({\bf G} +  {\bf k})^2  \equiv |{\bf G}|^2 + 2{\bf G} \cdot {\bf k} + |{\bf k}|^2$. Eigenvalue problem (\ref{Eig_Fourier_Form}) is solved {numerically}  for {all} ${\bf k}$ in the first Brillouin zone (BZ) given by $[-\pi ,\pi] \times [-\pi , \pi]$.  After solving (\ref{Eig_Fourier_Form}), the Fourier coefficients { $\widehat{u}_{p,mn}({\bf k})$} are returned to the series (\ref{U_Fourier})  and summed to give the Bloch wave. {The Bloch modes are normalized by fixing $\iint_{\rm UC} | E_p({\bf r}, {\bf k})|^2 \epsilon({\bf r}) \tilde{\mu}({\bf r})~  d{\bf r }= 1$, where $\text{UC}$ denotes the unit cell $[-1/2, 1/2] \times [-1/2 , 1/2].$}

 {The eigenvalue equation in (\ref{master_eqn_bloch_sub}) always contains spatial inversion symmetry since it is invariant under the transformations ${\bf k} \rightarrow - {\bf k}$ and ${\bf r} \rightarrow - {\bf r}$, and as a result the spectrum exhibits the inversion symmetry $\omega({\bf k}) = \omega(-{\bf k})$. {When} $\mathcal{M}({\bf r}) \not= {\bf 0}$ time reversal symmetry is broken{; i.e. the} imaginary term in (\ref{master_eqn_non_dim}) prevents the equation coming back to itself when ${\bf k} \rightarrow - {\bf k}$ and the complex conjugate of the equation is taken.}

First we numerically solve (\ref{Eig_Fourier_Form}) in the absence of any magnetic field. In Fig.~\ref{spectral_curves}(a) the first four dispersion bands are found to consist of two subsets: the (isolated) first band and the ({touching}) second through fourth bands. The band {touching} points occur at the origin ($\Gamma$) and corner ($M$) of the Brillouin zone. Next we consider {an} intermediate magnetized case that will play a crucial role in deriving our tight-binding system below. To obtain the bands in Fig.~\ref{spectral_curves}(b) we solve (\ref{master_eqn_bloch_sub}) with $\mathcal{M}({\bf r}) = {\bf 0}$, but $\tilde{\mu} \not= 0$ {on the right-hand side of the equation}. Again, the second, third, and fourth bands are all entangled, however this is still a system with time-reversal symmetry (and hence well-localized Wannier modes). We observe that our numerically obtained eigenfunctions $\widehat{u}_{mn}({\bf k})$ are real in this case.

Finally we compute the fully magnetized problem ($\mathcal{M}({\bf r} )\not= {\bf 0}$) and display the bands in Fig.~\ref{spectral_curves}(c). {Importantly, time-reversal} symmetry has been broken and a gap has opened between the second and third bands, as well as {between} the third and fourth bands. It is within these newly formed gaps  we later find topologically protected edge modes when a sharp edge boundary is introduced. The frequency gap ranges {between the second and third bands are} found to be $(0.525,0.571)$ and {between the third  and fourth bands are}   $(0.609,0.619)$. {{These values are in good agreement with \cite{wang1}.}  In dimensional units, a lattice constant of $a = 40$mm  means} these frequency gaps correspond to $(3.93,4.28)$ GHz and $(4.56,4.60)$ GHz, respectively. 

{{As stated above,} introducing the $\mathcal{M}({\bf r}) \not= {\bf 0}$ term in (\ref{master_eqn_non_dim}) breaks time-reversal symmetry{; in turn this} generates bands with nontrivial Chern number \cite{haldane1, haldane2}. {The} continuous Chern number of the $p^{\rm th}$ spectral band {is defined} by 
\begin{equation}
\label{chern_define}
C_p =   \frac{1}{2 \pi i} \iint_{\rm BZ} (\nabla_{\bf k} \times {\bf A}_p) \cdot \widehat{{\bf z}} ~ d{\bf k} ~ ,
\end{equation}
where ${\bf A}_p({\bf k}) = \langle u_p({\bf r}, {\bf k}) | \partial_{k_x} u_p({\bf r}, {\bf k})   \rangle_{{\rm UC},\epsilon \tilde{\mu} }~  \widehat{\bf x} + \langle u_p({\bf r}, {\bf k}) | \partial_{k_y} u_p({\bf r}, {\bf k})   \rangle_{{\rm UC},\epsilon \tilde{\mu}}   ~ \widehat{\bf y} $ is the Berry connection defined in terms of the complex inner product $$\langle f({\bf r}) | g({\bf r})  \rangle_{{\rm UC}, \epsilon \tilde{\mu}} = \iint_{\rm UC} f({\bf r})^* g({\bf r}) \epsilon({\bf r}) \tilde{\mu}({\bf r}) d{\bf r} .$$ The numerical algorithm in \cite{fukui} is used to compute (\ref{chern_define}). {As indicated in Fig.~\ref{spectral_curves}(c) the second and third bands 
acquire} a nontrivial Chern number{; which agrees with those found in  \cite{wang1}.} The bulk-edge correspondence indicates that topologically protected edge modes can be found in the band gaps when an edge boundary is introduced \cite{hatsugai}.  Topologically protected states are calculated {below} in Sec.~\ref{discrete_edge_bands_sec}; they are  consistent with the Chern numbers found here.}

{\section{A Perturbation Approach to  Wannier Expansion}} 
\label{wannier_sec}

A tight-binding model is {an approximation that 
transforms a PDE into} {a system of differential equations}. 
These approximations are most effective in {{deep} lattice regimes} where the Bloch mode can be expanded in terms of rapidly decaying basis functions {centered at different spatial locations. These basis functions are sometimes referred to as Wannier modes. Approximations to Wannier modes, termed orbitals, are also frequently used. The methodology is most useful when  the Wannier modes/orbitals are rapidly decaying.}
The overlap between {rapidly decaying Wannier modes/orbitals} is significant only among nearby {sites} 
and, as a result, the infinite series can be truncated to a manageable number of interactions. Typically the first few neighbors {(nearest/next-nearest)} are enough to gain a {good} approximation of the full problem.

The Bloch wave in (\ref{bloch_wave_define}) is periodic in ${\bf k}$ and {hence} can be expanded in terms of the Fourier series $E({\bf r}, {\bf k}) = \sum_{m,n} w_{m,n}({\bf r}) \exp(i {\bf R}_{mn} \cdot {\bf k})$. The Fourier coefficients are the Wannier modes and their decay rate in ${\bf r}$ 
 {depends on the} 
smoothness of $E({\bf r}, {\bf k})$ in ${\bf k}.$ When {the magnetic field
$\mathcal{M}({\bf r})$ in (\ref{master_eqn_non_dim}) is nonzero} 
time-reversal symmetry is lost and the system acquires a nontrivial Chern number \cite{wang1}. In this regime the Bloch function $E({\bf r}, {\bf k})$ is characterized by a {discontinuity} in ${\bf k}$ \cite{brouder}. 
As such, the Bloch wave does not have even one full derivative throughout the Brillouin zone and consequently the {direct} Wannier modes decay too slowly to be useful in a tight-binding model. 

Below we describe a perturbative approach to obtain a tight-binding model where the Wannier modes from a closely related 
 {equation}  are used to approximate the full 
 {problem.} These {approximate} 
 modes  {lead to 
 a discrete system of differential equations} that accurately describe the spectral bands. Moreover this discrete system supports a set of topologically protected modes that propagate unidirectionally around lattice defects {without backscatter}.

Let us motivate the perturbative approach. The essence of the method is to exploit the scales between the (relatively) weak $\mathcal{M}({\bf r})$ term in comparison to the (relatively) strong effective potential $\epsilon ({\bf r}) \tilde{\mu} ({\bf r})$ term in Eq.~(\ref{master_eqn_non_dim}). Inside the lattice rods the {amplitudes of the} inverse permeability tensor elements in (\ref{inverse_tensor}) are $\tilde{\mu}({\bf r}) \approx 3 $ and $\eta({\bf r}) \approx -0.29 $, while the amplitude of the effective potential  is $\epsilon \tilde{\mu} \approx 45$. For the smoothed potentials in (\ref{potential_model}) the {root mean squares of these magnetic terms over {one} 
unit cell are $\sqrt{\iint_{UC} |\partial_x \ln \tilde{\mu}|^2 d{\bf r}} = \sqrt{\iint_{UC} |\partial_y \ln \tilde{\mu}|^2 d{\bf r}} = 2.70$ and $\sqrt{\iint_{UC}  | \tilde{\mu} \eta_y |^2 d{\bf r}} = \sqrt{\iint_{UC}  | \tilde{\mu} \eta_x |^2 d{\bf r}} = 1.88$. In comparison to the strength of the effective permittivity term $\sqrt{\iint_{UC} |\epsilon({\bf r}) \tilde{\mu}({\bf r}) |^2 d{\bf r}} = 7.65$, the strength of these individual magnetic terms are $35\%$ and $25\%$, respectively.}  {In fact{,} it is the {(weaker)} second term that is responsible for the topological effects.} 
Relative to the effective potential $\tilde{\mu}({\bf r}) \epsilon ({\bf r})$ these terms are small, hence inclusion of the $\mathcal{M}({\bf r})$ term in (\ref{master_eqn_non_dim}) can be viewed as a perturbation to the same equation when $\mathcal{M}({\bf r}) = {\bf 0}$.  {While} this perturbation ratio is on the large side {we} find it does yield quite stable and accurate results.

\subsection{Fourier-Wannier Modes}
\label{wannier_modes_sec}

We now give a brief review of the Wannier functions and their properties. The spectrum of (\ref{master_eqn_non_dim}) consists of bulk dispersion surfaces{, or bands,} that are ordered by increasing frequency and are indexed by $p = 1,2,\dots${.} 
The full Bloch {wave} is written as a linear combination of the Bloch modes from the different bands i.e.
\begin{equation}
\label{bloch_wave_linear_comb}
E({\bf r},{\bf k}) = \sum_{p = 1}^{\infty} \alpha_p({\bf k}) E_p({\bf r},{\bf k}) ~ .
\end{equation}
{As mentioned earlier,} the Bloch function is periodic in ${\bf k}${. Hence} each Bloch mode is expanded in terms of a Fourier series 
\begin{equation}
\label{fourier_series}
E_p({\bf r},{\bf k}) =  \sum_{m,n = - \infty}^{\infty}   w_{p,mn}({\bf r}) e^{i {\bf k} \cdot {\bf R}_{mn}} ~ ,
\end{equation}
{where, under certain conditions, the Wannier coefficients are exponentially localized in space.}
The Wannier mode corresponding to the  $p^{\rm th}$ band is computed directly from the Bloch mode by 
\begin{equation}
\label{wannier_define}
w_{p,mn}({\bf r}) = \frac{1}{4 \pi^2} \iint_{\rm BZ} e^{- i {\bf k} \cdot {\bf R}_{mn}} E_p({\bf r}, {\bf k}) ~  d{\bf k} ~,
\end{equation}
where $4 \pi^2$ is the area of the first Brillouin zone (BZ).   {The Wannier functions} decay to zero as $|{\bf r}| \rightarrow \infty$.  {A key property of Wannier functions is that in a specific band they are related through {integer translations}. 
This implies
\begin{equation}
\label{wannier_shift}
w_{p,mn}({\bf r}) =  w_{p,00}({\bf r} - {\bf R}_{mn}) ~, ~~~~ R_{mn}= m{\bf e}_1+n {\bf e}_2~ .
\end{equation}
Hence the electric field is being expanded in terms of an infinite number of copies of the same Wannier mode located {at} 
 each lattice site ${\bf R}_{mn}$.  
{We} define the complex inner product 
\begin{equation}
\langle f({\bf r}) | g({\bf r})  \rangle_{\mathbb{R}^2} = \iint_{\mathbb{R}^2} f({\bf r})^* g({\bf r})~ d{\bf r} ~ .
\end{equation}
 The Wannier functions form an orthogonal basis on $\mathbb{R}^2$ and satisfy the following orthogonality {relation}
\begin{equation}
\label{wannier_orthogonality}
 \langle w_{\ell,m'n'} |  \epsilon \tilde{\mu} w_{p,mn} \rangle_{\mathbb{R}^2} = \delta_{mm'} \delta_{nn'} \delta_{\ell p} ~ ,
\end{equation}
where $\delta_{ij}$ is the Kronecker delta.

Due {to} the nature in which eigenvalue problem (\ref{master_eqn_bloch_sub}) is solved numerically there is non-uniqueness in the Bloch modes which causes discontinuities in ${\bf k}$. Since the eigenvalue problem is linear, at each value of ${\bf k}$ the {normalized}  eigenfunction is only known up to an arbitrary constant i.e. $ C({\bf k}) E({\bf r}, {\bf k})$. In general, the function {$C({\bf k})$} is not {smooth which means the corresponding Wannier mode decays slowly.} 
To} improve the localization properties of the mode{s from bands $p = 2,3,4$} we utilize the Marzari and Vanderbilt (MV) algorith{m} 
\cite{marzari}. {This} algorithm applies a gradient descent method to find Wannier modes that minimize  the spread or variance. 
Details of our implementation of the MV algorithm are given in Appendix~\ref{MV_alg_sec}. {The first band is isolated from all others in Fig.~\ref{spectral_curves} and the corresponding Bloch mode is smoothed by the rescaling method described in Appendix~\ref{initial_guess_sec}. }

{The numerically computed Wannier modes are shown in Fig.~\ref{wannier_modes_plot}. Recall that these modes are calculated from the $\mathcal{M}({\bf r})= {\bf 0}$ problem in (\ref{master_eqn_non_dim}). As a result, this equation has time-reversal symmetry and the corresponding Wannier modes are real \cite{marzari}. The first, third, and fourth Wannier modes are found to be centered at the origin, whereas the second mode is centered at the off-rod{, half-integer} point $\left(-\frac{1}{2} , -\frac{1}{2} \right).$ This somewhat surprising result that the $p = 2$ Wannier mode is centered off-site is a consequence of the optimization procedure used in the MV algorithm ({see also \cite{Busch}).}  As a final note, depending on the initial guess, sometimes these Wannier modes can appear in a different order. If this occurs, we re-order them so that off-site mode shown in {Fig.~\ref{wannier_modes_plot}(b) corresponds to $p = 2.$}

\begin{figure} [h]
\centering
\includegraphics[scale=0.42]{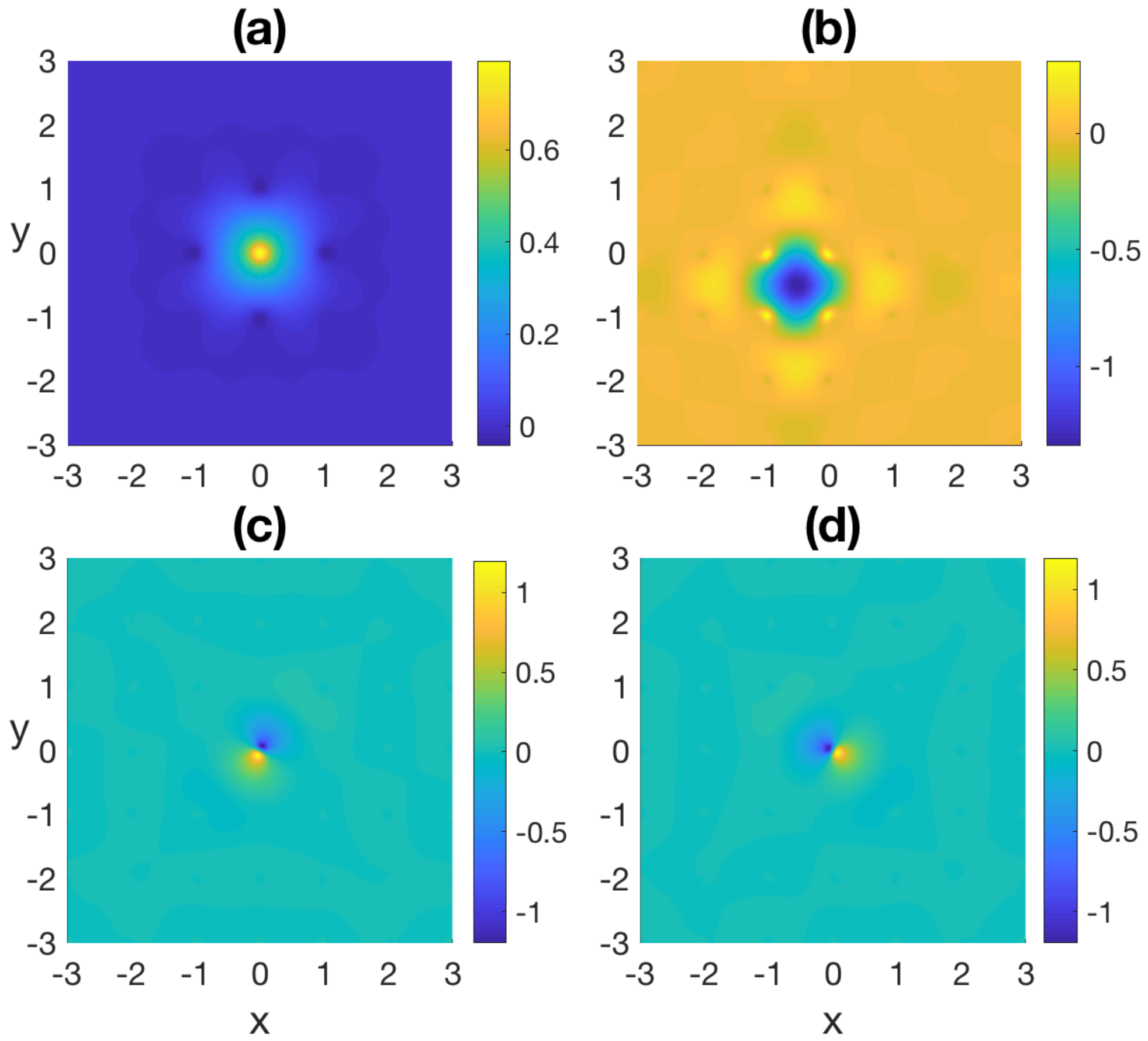}
\caption{{Plot of Wannier modes (all from (6) with $\mathcal{M}({\bf r}) = {\bf 0}$): (a) $w_{1,00}({\bf r})$, (b) $w_{2,00}({\bf r})$, (c) $w_{3,00}({\bf r})$, and (d) $w_{4,00}({\bf r})$.}}
\label{wannier_modes_plot}
\end{figure}

 {The tight-binding approximation used below is most effective when the Bloch function is expressed in terms of a rapidly decaying set of basis functions. Ideally these functions decay exponentially{ fast} in space. Side profiles of the Wannier modes in Fig.~\ref{wannier_modes_plot} are shown in Fig.~\ref{wannier_profiles_plot}. In each case a curve is selected in order highlight the rapid decay of the Wannier modes. {Numerically,} all cases are found to decay exponentially fast, however the $p=2$ mode in Fig.~\ref{wannier_profiles_plot}(b) {is the slowest of the four.} 
 This is expected based on the spread minimization results in Appendix~\ref{MV_converge_sec}.}

\begin{figure} [h]
\centering
\includegraphics[scale=0.48]{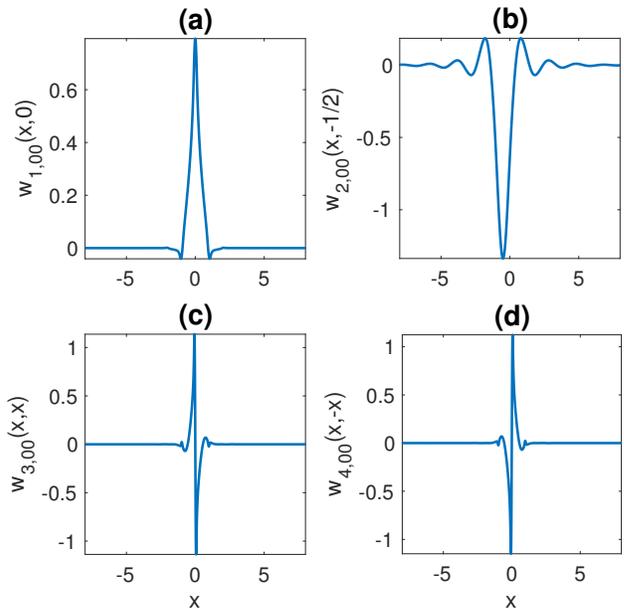}
\caption{{Profiles of Wannier modes from Fig.~\ref{wannier_modes_plot}, respectively. The Wannier modes are shown along the lines: (a) $y = 0$, (b) $ y = -1/2$, (c) $y = x$, and (d) $y = -x.$}}
\label{wannier_profiles_plot}
\end{figure}

With the key assumption{, about how to find Wannier modes,} {addressed} we {next provide an outline} 
of how we obtain our discrete model: 
\begin{enumerate}
\item Numerically compute the Bloch functions (see Sec.~\ref{num_comp_spec_bands}).
\item Smooth Bloch functions though gradient descent optimization (see Appendix~\ref{MV_alg_sec}).
\item Compute corresponding Wannier functions (see Fig.~\ref{wannier_modes_plot}).
\item Use Wannier modes to calculate the coefficients of {the} discrete model (see Sec.~\ref{TbA_sec}).
\item Solve resulting tight-binding approximation: coupled system of ODEs (see Secs.~\ref{spec_band_sec} and \ref{time_evolve_sec}).
\end{enumerate}

\section{Derivation of Tight-binding  Model}
\label{TbA_sec}

We now derive our discrete tight-binding model using the Wannier functions found above as a basis. Combine the electric field given in (\ref{bloch_wave_linear_comb})-(\ref{fourier_series}) in terms of the {Wannier } expansion
\begin{equation}
\label{edge_mode_expand}
E({\bf r},{\bf k}) =  \sum_{p = 1}^{\infty}     \sum_{m,n = - \infty}^{\infty} \alpha_{p,mn}({\bf k})  w_{p,mn}({\bf r}) ~ ,
\end{equation}
 {where $\alpha_{p,mn}$ combines the coefficient in (\ref{bloch_wave_linear_comb}) and the phase of (\ref{fourier_series}).} The $w_{p,00}({\bf r})$ Wannier modes are shown in Fig.~\ref{wannier_modes_plot}. All other Wannier modes are obtained through {integer} {shifts, see} {Eq.~}(\ref{wannier_shift}). Take expansion (\ref{edge_mode_expand}) and substitute it into the {\it full} $(\mathcal{M}({\bf r}) \not= {\bf 0})$ system (\ref{master_eqn_non_dim}) to obtain
\begin{align}
\nonumber
   \sum_p    \sum_{m,n} \alpha_{p,mn} & \bigg[  -   \nabla^2 w_{p,mn}  +   \mathcal{ M}({\bf r}) \cdot \nabla w_{p,mn} \bigg] \\
& = \omega^2 ({\bf k})  \sum_p     \sum_{m,n} \alpha_{p,mn} \epsilon \tilde{\mu}   w_{p,mn} ~ .
\end{align}
Multiplying through by $w^*_{\ell,m'n'}$ and integrating over $\mathbb{R}^2$ yields
\begin{align}
 & \sum_p     \sum_{m,n} \alpha_{p,mn}    \bigg[ \langle \partial_x w_{\ell,m'n'} | \partial_x w_{p,mn} \rangle_{\mathbb{R}^2} \\ \nonumber
 &  + \langle \partial_y w_{\ell,m'n'} | \partial_y w_{p,mn} \rangle_{\mathbb{R}^2}+ \langle w_{\ell,m'n'} | \mathcal{M}({\bf r}) \cdot \nabla w_{p,mn}   \rangle_{\mathbb{R}^2} \bigg] \\ \nonumber
&= \omega^2({\bf k})  \sum_p     \sum_{m,n} \alpha_{p,mn} \langle w_{\ell,m'n'} |  \epsilon \tilde{\mu} w_{p,mn} \rangle_{\mathbb{R}^2} ~ , 
\end{align}
where integration-by-parts  has been applied. 
Utilizing the Wannier orthonormality in (\ref{wannier_orthogonality}), the right-hand side reduces to
\begin{align}
 & \sum_p     \sum_{m,n} \alpha_{p,mn}    \bigg[ \langle \partial_x w_{\ell,m'n'} | \partial_x w_{p,mn} \rangle_{\mathbb{R}^2} \\ \nonumber
 &  + \langle \partial_y w_{\ell,m'n'} | \partial_y w_{p,mn} \rangle_{\mathbb{R}^2}+ \langle w_{\ell,m'n'} | \mathcal{M}({\bf r}) \cdot \nabla w_{p,mn}   \rangle_{\mathbb{R}^2} \bigg] \\ \nonumber
&= \omega^2({\bf k})  \alpha_{\ell,m'n'}  ~ .
\end{align}
For convenience let us rewrite $m = m' + \mu$ and $n = n' + \nu$, where $\mu , \nu \in \mathbb{Z}$. So the above equation becomes
\begin{align}
\label{int_eqn_111}
 \sum_p     \sum_{\mu,\nu} & \alpha_{p,m'+\mu,n'+\nu}    \bigg[ \langle \partial_x w_{\ell,m'n'} | \partial_x w_{p,m'+\mu,n'+\nu} \rangle_{\mathbb{R}^2} \\ \nonumber
 & + \langle \partial_y w_{\ell,m'n'} | \partial_y w_{p,m'+\mu,n'+\nu} \rangle_{\mathbb{R}^2} \\ \nonumber
 &+ \langle w_{\ell,m'n'} | \mathcal{M}({\bf r}) \cdot \nabla w_{p,m'+\mu,n+\nu}   \rangle_{\mathbb{R}^2} \bigg]  \\ \nonumber
 &= \omega^2 ({\bf k}) \alpha_{\ell,m'n'} ~ .
\end{align}
Now define the coefficients in terms of the Laplacian and magnetic terms, respectively, of
 \begin{align}
 \label{discrete_lap_coeff}
& \mathbb{L}^{\ell p}_{\mu\nu} \equiv \langle \partial_x w_{\ell,00} | \partial_x w_{p,\mu\nu} \rangle_{\mathbb{R}^2}  + \langle \partial_y w_{\ell,00} | \partial_y w_{p,\mu\nu} \rangle_{\mathbb{R}^2}  ~ ,\\
 \label{discrete_mag_coeff}
& \mathbb{M}^{\ell p}_{\mu\nu} \equiv \langle w_{\ell,00} | \mathcal{ M}({\bf r}) \cdot \nabla w_{p,\mu\nu} \rangle_{\mathbb{R}^2} ~.
\end{align}
Note that  all Wannier modes, for a given band, are merely translations of the $(m,n) = (0,0)$ Wannier mode.
The equations in (\ref{int_eqn_111}) are now rewritten as
\begin{equation}
\label{sum_infinite}
 \sum_{ p = 1}^{\infty}     \sum_{\mu,\nu = -\infty}^{\infty}   \left[ \mathbb{L}^{\ell p}_{\mu\nu}  + \mathbb{M}^{\ell p}_{\mu\nu} \right]  \alpha_{p,m+\mu,n+\nu} 
= \omega^2 ({\bf k}) \alpha_{\ell,mn} ~ ,
\end{equation}
where{, for convenience,} the primed index notation has been discontinue{d.} {Equation (\ref{sum_infinite}) is an algebraic eigenvalue problem for $\omega^2$ {as a function of ${\bf k}$}; later when we need to understand dynamics, this system is transformed to differential equations by replacing $\omega$ with $-i\partial_t$.}

The lattice sites of the tight-binding model consist of two interpenetrating square lattices: integer ($p = 1,3,4$) and half-integer ($p = 2$), centered, respectively, at the points
\begin{align}
\label{lattice_site_location_sets}
& W_i \equiv \left\{ (x_m , y_n)  \big| x_m = m , y_n = n \right\} ~ , \\ \nonumber
&W_h \equiv \left\{ (x_m , y_n)  \big| x_m = m-1/2 , y_n = n-1/2   \right\} ~ ,
\end{align}
where $m, n \in  \mathbb{Z}$. This means the $(m,n)$ {Wannier} mode for $p = 1,3,4$ $ [p=2]$ is centered at {the spatial location} $(m,n)$ $[(m -1/2, n - 1/2)]$. {Since the algorithm in \cite{marzari} mixes modes $p=2,3,4$ we cannot ascribe significance to any one of these modes.}
{As discussed above,} the Wannier functions in Fig.~\ref{wannier_modes_plot} are exponentially decaying functions. As such,  the coefficients in (\ref{discrete_lap_coeff}) and (\ref{discrete_mag_coeff}) are negligibly small {for $\sqrt{ \mu^2 +  \nu^2} \gg 1$}. The infinite series in (\ref{sum_infinite}) can be truncated to a tractable number of interactions by neglecting long-range interactions. This is the tight-binding approximation.

For the results in Secs.~\ref{spec_band_sec} and \ref{time_evolve_sec} we consider all neighboring sites that are distance of {one} or less away from a central lattice site. This amounts to a total of {17} ($p=2$) or {19} ($p = 1,3,4$)  interactions to consider and has been demonstrated to give reasonable results for a manageable number of interactions.  As the number of interactions increases the accuracy of the band diagrams tends to improve. In Appendix~\ref{asym_limit_tba} we present a few band diagrams {that have additional} 
interactions and see how they compare.

\begin{figure} [h]
\centering
\includegraphics[scale=0.8]{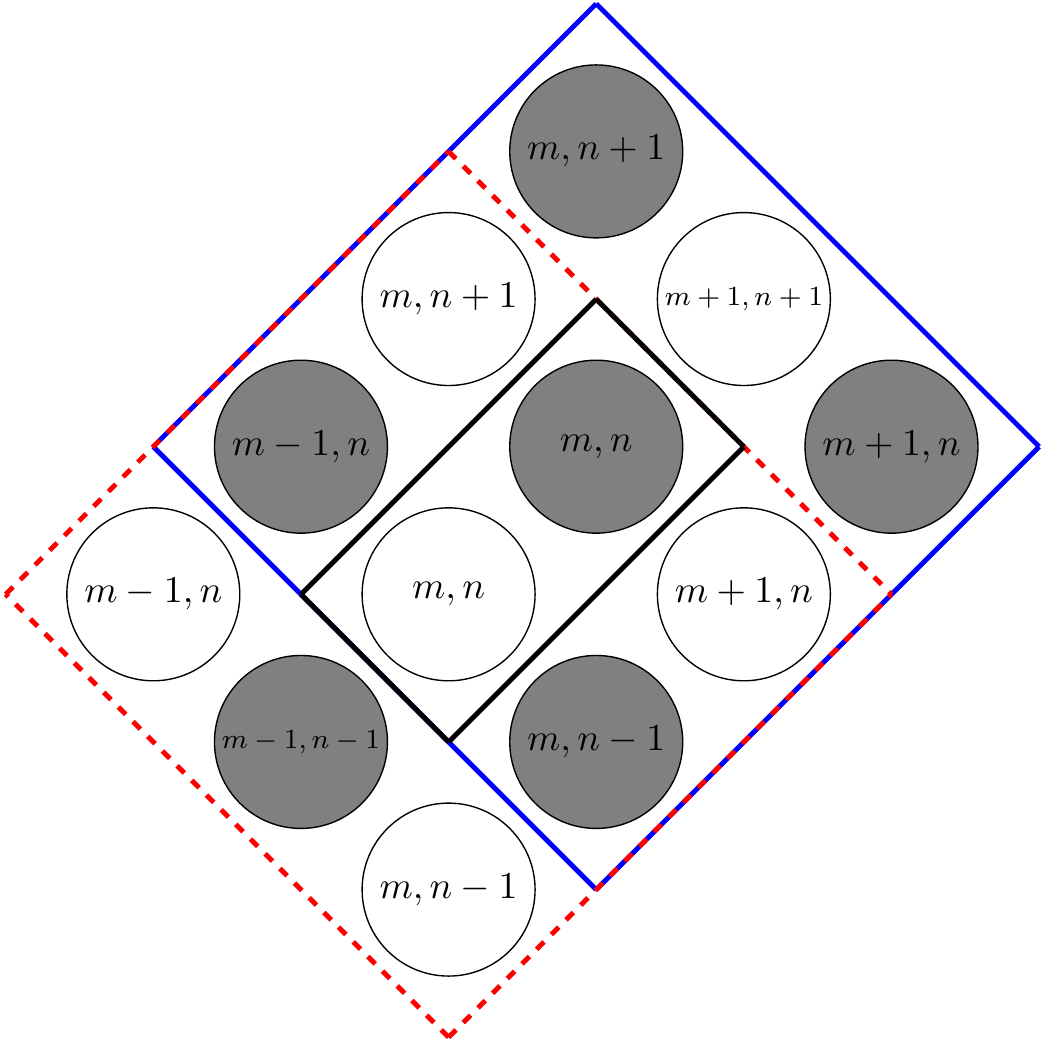}
\caption{(color online) Wannier mode locations and tight-binding interactions. Gray circles indicate the location of the  $p = 1,3,4$  Wannier modes {(integer sites)}. White circles are {where the $p = 2$ modes are centered (half-integer sites)}. For the gray (white) circle in the center black box, the solid blue (dashed red) box encloses all Wannier {interactions of distance one or less {away}.}}
\label{draw_TB_interact_labeled}
\end{figure}

{A diagram organizing the interactions that are taken into account in the tight-binding approximation is displayed in Fig.~\ref{draw_TB_interact_labeled}. The circles correspond to locations of the Wannier modes given in (\ref{lattice_site_location_sets}): white circles at half-integer sites  and gray circles at integer sites. Recall the $p = 2$ mode is centered at (white) half-integer sites [see Fig.~\ref{wannier_modes_plot}(b)], while the $ p = 1,3,4$ modes are centered at (gray) integer sites [see Figs.~\ref{wannier_modes_plot}(a,c-d)]. Define ${\bf d}_s$ to be the set of all lattice sites 
that are  euclidean distance $s$ or less away. The $s = 1$ case is {shown} in Fig.~\ref{draw_TB_interact_labeled}. For the central white (gray) lattice site in the black box, the red (blue) 
 box indicates  
all Wannier mode locations that are  distance $s$ or less away. These are the Wannier modes whose interactions  will be considered. All other  interactions with Wannier modes outside ${\bf d}_{s}$ are neglected. 

For example, in the ${\bf d}_1$ case modes centered at the central white (gray) circle interact with itself and modes located at the 
four nearest gray (white) circles, which are a distance of $1/\sqrt{2}$ away. Additionally, the central white (gray) site interacts with 
the four next-nearest white {(gray)} circles, distance $1$ away.}  

The infinite system in (\ref{sum_infinite}) is truncated to the four lowest spectral bands and rewritten as  {
\begin{equation}
\label{tb_eqn_freq}
 \sum_{p= 1}^{4}    \sum_{\mu,\nu \in {\bf d}_s}     \left[ \mathbb{L}^{\ell p}_{\mu\nu}  + \mathbb{M}^{\ell p}_{\mu\nu} \right] \alpha_{p,m+\mu,n+\nu}
= \omega^2({\bf k})  \alpha_{\ell,mn} ~ , 
\end{equation}
for $\ell = 1,2,3,4,$.} 
Taking the inverse Fourier transform of this equation is equivalent to replacing $\omega$ with  $- i \partial_t$. Hence the time-dependent coupled mode equations are given by {
\begin{equation}
\label{tb_eqn_time}
\frac{d^2 a_{\ell,mn}}{d t^2} +  \sum_{p = 1}^4     \sum_{\mu,\nu \in {\bf d}_s}    \left[ \mathbb{L}^{\ell p}_{\mu\nu}  + \mathbb{M}^{\ell p}_{\mu\nu} \right] 
 a_{p,m+\mu,n+\nu}  = 0 ~ ,
\end{equation}
 } where $ \ell = 1,2,3,4$ and $a_{p,mn}(t)  = \alpha_{p,mn}\exp(i \omega t).$ {The updated (time-dependent) electric field is given by
 \begin{equation}
 \label{e_field_time_dep}
 \overline{E}({\bf r}, {\bf k},t) =  \sum_{p }     \sum_{m,n} a_{p,mn}({\bf k},t)  w_{p,mn}({\bf r}) ~ .
 \end{equation}
 Adding the complex conjugate to this function gives the (real) field defined in (\ref{maxwell}), i.e. ${\bf E}({\bf r},t) = (0,0, \overline{E}({\bf r},t)) +  ( 0 , 0, \overline{E}^*({\bf r},t)).$ {The function in (\ref{e_field_time_dep})} 
 is effectively the inverse Fourier transform of the expansion in (\ref{edge_mode_expand}).}
 
 It is from Eqs.~(\ref{tb_eqn_freq}) and (\ref{tb_eqn_time}) we derive our discrete results {which are discussed} below. In Sec.~\ref{spec_band_sec} we compute the bulk and edge bands directly from (\ref{tb_eqn_freq}). The bands illuminate the relationship between the presence of topological invariants and the corresponding number and nature of gapless edge states. 
 {Then in} Sec.~\ref{time_evolve_sec} we solve the  initial {boundary} {value} problem in (\ref{tb_eqn_time}) and observe the unidirectional propagation of a topologically protected mode.

\section{Spectral Bands: Discrete Model}
\label{spec_band_sec}

In this section we compute discrete approximations of the spectral bands  for the bulk (infinite) and edge (semi-infinite) problems.  For the bulk problem we can compare directly  with the spectral bands found in Sec.~\ref{num_comp_spec_bands}. From our discrete approximation of the bands we establish the presence of nontrivial Chern numbers in the model. {Nonzero Chern numbers indicates the presence of topologically protected gap modes when an edge is introduced, via the {bulk-edge} correspondence \cite{hatsugai}. } 
Next, in the edge problem we {consider {a semi-infinite strip domain with Dirichlet zero boundary conditions along the top and bottom sides. Beyond these boundaries (outside the lattice region) the electric field} 
is {assumed to be} negligibly small. {Perpendicular to} these walls we  look for localized eigenmodes called {\it edge modes}. Topologically protected modes manifest themselves as a family of localized eigenmodes whose corresponding eigenvalues span the gap between two bulk bands.

To obtain the coefficients used {in} this paper, a {basic} simulation that ran  
 {on} a standard desktop {computer} {was used.} 
 {Most of that time was spent on computing the Bloch modes in (\ref{Eig_Fourier_Form}) 
and the discrete model coefficients in (\ref{discrete_lap_coeff}) and (\ref{discrete_mag_coeff}). }
 {This} is a one-time overhead cost. 
With these in hand, the discrete system in this section can be solved 
 {on} a laptop computer.

\subsection{Discrete Approximation of Bulk {Bands}} 
\label{discrete_disperse_surf}

To approximate the spectral dispersion surfaces found in Sec.~\ref{num_comp_spec_bands} we look for plane wave solutions of the form
$ \alpha_{p,mn}({\bf k}) = \alpha_{p}({\bf k}) e^{ i {\bf k} \cdot {\bf R}_{mn} } $
in Eq.~(\ref{tb_eqn_freq}). 
The governing coupled-mode system  is {
\begin{equation}
\label{tb_eig_sys}
 \sum_{p=1}^4     \sum_{\mu,\nu \in {\bf d}_s}    \left[ \mathbb{L}^{\ell p}_{\mu\nu}  + \mathbb{M}^{\ell p}_{\mu\nu } \right] 
 \alpha_{p} e^{i {\bf k} \cdot {\bf R}_{\mu\nu}}   = \omega^2({\bf k}) \alpha_{\ell}({\bf k}) ~ , 
\end{equation}
} for $\ell = 1,2,3,4$. This system can be rewritten as the $4 \times 4$ eigenvalue problem
\begin{equation}
\label{tb_eig_sys_matrix}
\mathbb{N}({\bf k}) {\bf a} = \omega^2({\bf k}) {\bf a} ~ , ~~~~ {\bf a}({\bf k}) = \begin{pmatrix} \alpha_1 , \alpha_2,  \alpha_3, \alpha_4 \end{pmatrix}^T ~ ,
\end{equation}
with matrix elements {
\begin{equation*}
\left[ \mathbb{N}_{\ell p}({\bf k}) \right] =  \sum_{\mu,\nu \in {\bf d}_s}  \left[ \mathbb{L}^{\ell p}_{\mu\nu}  + \mathbb{M}^{\ell p}_{\mu\nu} \right] e^{i {\bf k} \cdot {\bf R}_{\mu\nu}} ~ . 
\end{equation*}
} Eigenvalues{/eigenfunctions} for {(\ref{tb_eig_sys_matrix})} are computed for values of ${\bf k}$ along the perimeter of the irreducible Brillouin zone [see Fig.~\ref{sq_lattice_plot}(c)]. {In all examples considered below we take $s = 1$ i.e. interactions between nearest and next-nearest neighbor (see Fig.~\ref{draw_TB_interact_labeled}).} In Fig.~\ref{MO_spec_bands_compare} the discrete {approximations} (circles)  
are plotted on top of the numerically generated bands {(solid curves)}{,} originally shown in {Figs.~\ref{spectral_curves}(b) and \ref{spectral_curves}(c)}. In all cases we only show the real part of $\omega$, as the imaginary part is typically quite small. First, consider when $\mathcal{M}({\bf r}) = {\bf 0}$ (equivalently $\mathbb{M}^{\ell p}_{\mu \nu} = 0$). The top three bands of Fig.~\ref{MO_spec_bands_compare}{(a)} form a disjoint subset of entangled bands that touch at the $\Gamma(0,0)$ and $M(\pi,\pi)$ points.

\begin{figure} [h]
\centering
\includegraphics[scale=0.35]{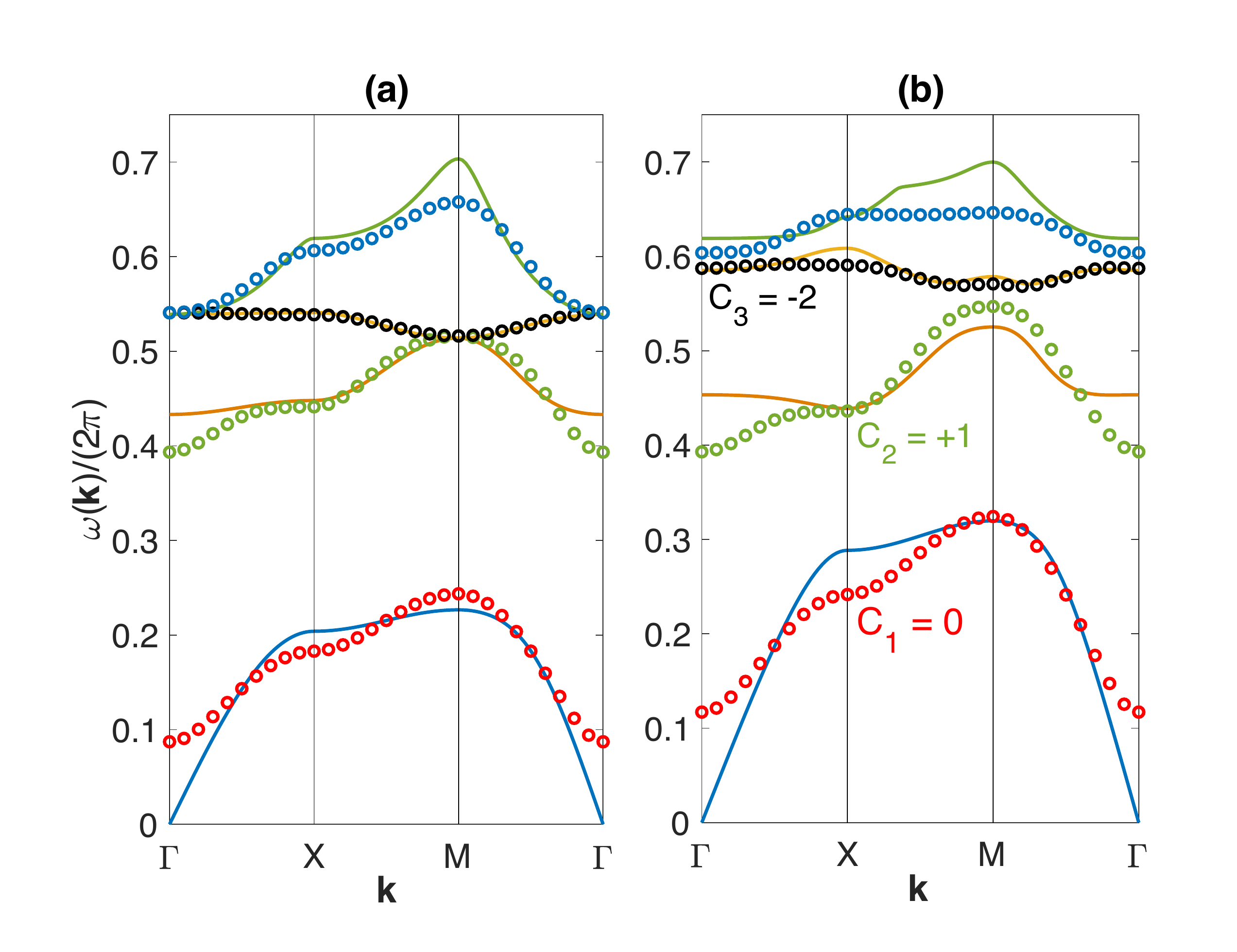}
\caption{Comparison between discrete spectral bands and the numerical bands shown in Figs.~\ref{spectral_curves}(b-c). (a) $\mathbb{M}^{\ell p}_{\mu \nu}= 0$ ($\mathcal{M}({\bf r}) = {\bf 0}$) and (b) $\mathbb{M}^{\ell p}_{\mu \nu} \not= 0 $ ($\mathcal{M}({\bf r}) \not= {\bf 0}$). The numerical  bands are denoted by solid curves and the discrete by circles. 
Included are the Chern numbers, defined in (\ref{chern_define}), computed from the discrete model.}
\label{MO_spec_bands_compare}
\end{figure}

Now we turn our attention to the  $\mathcal{M}({\bf r}) \not= {\bf 0}$ (equivalently $\mathbb{M}^{\ell p}_{\mu \nu} \not= 0$) case.  Physically, this corresponds to the introduction of an external magnetic field and breaking {of} time-reversal symmetry. Overall the discrete bands shown in Fig.~\ref{MO_spec_bands_compare}(b) serve as good approximations to the {numerics}. Among the {upper three} bands the relative error is {11.5\%} or less across the {entire} Brillouin zone.   Moreover these bulk bands also preserve the inversion symmetry of the lattice i.e. $\omega({\bf k}) = \omega(-{\bf k})$.  A quantitative comparison between the bands is summarized in Table~\ref{num_wannier_modes_comp} of Appendix~\ref{asym_limit_tba} {for different values of $s$}.

We note a weak instability  in {$\mathcal{M}({\bf r}) \not= {\bf 0}$} case for some of the {frequencies}, with $\Im \omega({\bf k})$ on the order of 
 {$0.01$} or less. We attribute this spurious instability to the perturbation approximation underlying our approach. In spite of this we point out that the edge modes shown in the next section are quite stable, the unstable modes appear to only occur in bulk modes. As long as we 
 {do} not excite any unstable frequencies, {we had no problem computing over long times.} Finally, we point out that there is no instability in the $\mathcal{M}({\bf r}) = {\bf 0}$ case. 

Inspection of the discrete bands in Fig.~\ref{MO_spec_bands_compare}(b) reveals the formation of two gaps: one between the second and third bands and another between the third and fourth bands. There is a full gap between the second and third bands throughout the entire Brillouin zone, which we refer to as a {\it zone gap}. 
 There is also small zone gap that exists between the third and fourth bands. The frequency gaps are in the ranges {$ (0.547 , 0.568)$ and $ (0.592 , 0.604)$, respectively.}

{As indicated above,} breaking time-reversal symmetry is associated with nontrivial Chern numbers in the spectral bands  \cite{haldane1,haldane2}. The Chern number is a time-independent topological invariant that is robust against lattice defects and deformations {along the boundary}. Moreover, through the bulk-edge correspondence, a nonzero Chern number indicates the presence of a topologically protected edge state \cite{hatsugai}. The 2D Chern number of the $p^{\rm th}$ spectral band is defined by (\ref{chern_define}) where {
\begin{equation}
{\bf A}_p({\bf k}) = \langle {\bf a}_p | \partial_{k_x} {\bf a}_p   \rangle  \widehat{\bf x} + \langle {\bf a}_p | \partial_{k_y} {\bf a}_p   \rangle \widehat{\bf y} ~ ,
\end{equation}
} is the Berry connection given in terms of the eigenmodes of (\ref{tb_eig_sys_matrix}). {The vector ${\bf a}_p$ is an eigenvector of the matrix (\ref{tb_eig_sys_matrix}) corresponding to the $p^{\rm th}$ eigenvalue, sorted in ascending order.}  Here we take the complex inner-product  $\langle {\bf f}({\bf k}) | {\bf g}({\bf k}) \rangle \equiv {\bf f}({\bf k})^{\dag} {\bf g}({\bf k})$ {for the vectors {\bf f} and {\bf g}}.  The Chern numbers are computed using the algorithm given in \cite{fukui}.  The {second and third}  bands in Fig.~\ref{MO_spec_bands_compare}(c) are found to possess nontrivial Chern numbers. {The Chern values for the first three bands agree with those found in the continuous problem, shown in Fig.~\ref{spectral_curves}(c).} 
  In the next section we show that the number and chirality of edge states is consistent with these Chern numbers.

\subsection{Discrete Edge  Bands}
\label{discrete_edge_bands_sec}

We now look for eigenmodes in the edge problem. This problem is formulated as the solution of (\ref{tb_eqn_freq}) on a strip domain that is {infinite}  in the $x(m)$-direction and finite with {Dirichlet zero} 
boundary conditions in the $y(n)$-direction. {Localized} along both the top and bottom walls we identify edge modes whose corresponding eigenvalues lie in the band gaps. Edge modes are eigenmodes characterized by exponential decay away from the  boundary wall. Furthermore, we {can} relate the quantity and chirality of these edge states to the Chern numbers found above.

Let us consider modes of the form
$ a_{p,mn}(t) = a_{p,n}(k_x) e^{ i (m k_x  + \omega t) }  $
 in  Eq.~(\ref{tb_eqn_time}). {Then the governing systems of equations is {
\begin{equation}
\label{tb_eqn_spec_edge2}
 \sum_{p = 1}^4 \sum_{\mu,\nu \in {\bf d}_s}    \left[ \mathbb{L}^{\ell p }_{\mu\nu}  + \mathbb{M}^{\ell p}_{\mu\nu} \right] 
 a_{p,n+\nu} e^{i \mu k_x}  = \omega^2(k_x) a_{\ell,n} ~ ,
\end{equation}
} for $\ell = 1,2,3,4.$ This eigenvalue problem {for $\omega^2$} is solved for values of $k_x$ in the interval $[-\pi ,\pi].$}

Now let us incorporate the top and bottom wall boundary conditions. An edge mode consists of a sharp termination at a boundary wall with exponential decay {perpendicular to} 
the wall, {which can be imposed by Dirichlet }  
zero conditions. As a result of the integer and half-integer sites, there are at least two different ways to set the edge. Here {we} concentrate on boundary walls whose top and bottom rows consist of integer lattice sites {i.e. of ferrite rods}. Recall these are the locations of the $p=1,3,4$ Wannier modes. This means we create a sharp termination of the problem immediately after the first or last integer lattice site rows (see Fig.~\ref{draw_TB_BC}(a) for an illustatration).
The Dirichlet zero boundary conditions we impose on {Eq.~(\ref{tb_eqn_spec_edge2}) {are}}
\begin{align}
\label{zero_BCs}
 p = 1,3,4: ~~~~~~ &a_{p,n}  = 0,  ~~~~ {\rm for }~~ |n| > N ~ , \\ \nonumber
 p = 2: ~~~~~~ &a_{p,n}  = 0,  ~~~~ {\rm for }~~ n \le -N ~ \text{and} ~ n > N ~ .
\end{align}
Notice there is one less row of half-integer lattice sites. The edge bands for half-integer boundary conditions are discussed in Appendix~\ref{diff_BC_sec}. Changing the boundary condition is found to affect the edge state eigenmode curves, but not the bulk bands.

The {result of solving (\ref{tb_eqn_spec_edge2}) with (\ref{zero_BCs})} for interactions of distance {one} or less is shown in Fig.~\ref{edge_bands_plot}. In both figures the solid black regions correspond to non-localized bulk modes, while the highlighted curves are gapless bands whose corresponding eigenmodes are localized {near the edges}. First, in Fig.~\ref{edge_bands_plot}(a) we show the dispersion bands that result when $\mathcal{M}({\bf r}) = {\bf 0}$. Here we see only bulk bands and no edge modes. Next, in Fig.~\ref{edge_bands_plot}(b) we set $\mathcal{M}({\bf r}) \not= {\bf 0}$ and see the opening of new band gaps is accompanied by a family of curves that span these gaps.

\begin{figure} [h]
\centering
\includegraphics[scale=0.38]{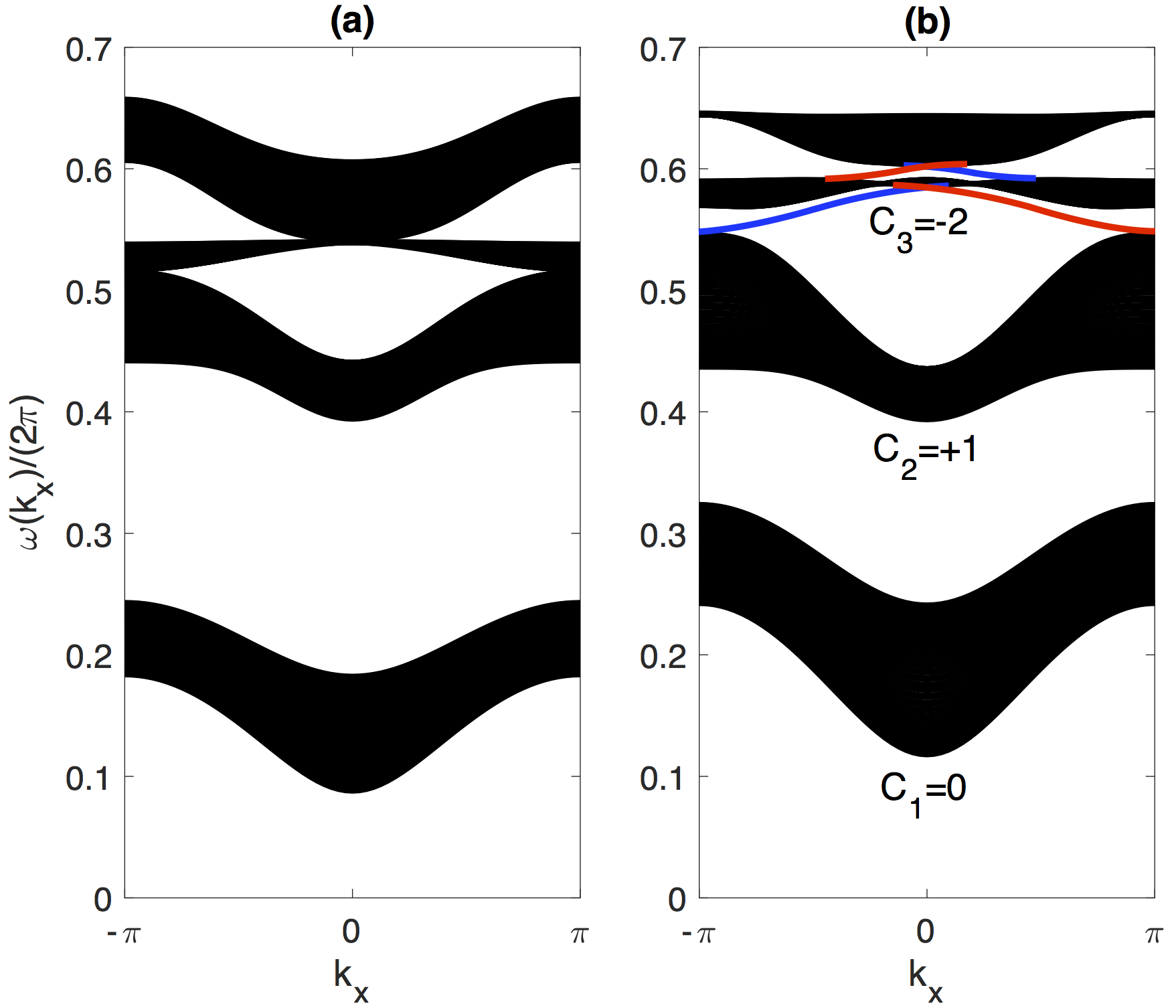}
\caption{The $p = 1,2,3,4$ dispersion bands {(ascending order)} found by solving edge problem (\ref{tb_eqn_spec_edge2}) {with (\ref{zero_BCs})} for (a) $\mathbb{M}^{\ell p}_{\mu \nu} = {0}$ and (b) $\mathbb{M}^{\ell p}_{\mu \nu} \not= {0}$.}
\label{edge_bands_plot}
\end{figure}

Next focus on the gapless edge bands that lie in-between the second, third and fourth bands. A closer view of the bands from Fig.~\ref{edge_bands_plot}(b) is shown in Fig.~\ref{edge_modes_position}(a). Each gap has a pair of edge modes, one on {the top boundary} (red curve) and one on {the bottom boundary} (blue curve), that constitute a single chiral state. The modes spanning the gap between the second and third bands have a clockwise orientation (as viewed from the positive $z$ direction) since a wide envelope along the bottom (top) edge will have negative (positive) group velocity. Note that we have taken solutions of the form $\exp(i (m k_x + \omega t))$, so the group velocity moves in the $\text{sgn} (- \omega'(k_x))$ direction. Two representative edge states are displayed in Figs.~\ref{edge_modes_position}(d) and \ref{edge_modes_position}(e) corresponding to the bottom and top walls, respectively. In the {eigenmodes} we see that {the (half-integer site) $p = 2$ mode has a much} smaller {magnitude} than the {(integer site)} $p = 3,4$ modes. This says that the electric field has a higher intensity close to the YIG rods. 

{We {also} notice that the $p =1$ {part of the} eigenmodes we find are effectively zero for large $\omega$ corresponding to the upper three bands, and vice versa for eigenfrequencies in the first bulk band. {There} is an apparent decoupling between the first mode and the second through fourth modes. This suggests that the system (\ref{tb_eqn_spec_edge2}) could be decoupled into two parts: $p = 1$ and $p=2,3,4$ and solved separately. We have implemented this decoupling and noticed that it did not significantly affect any of the band diagrams.}

\begin{figure} [h]
\centering
\includegraphics[scale=0.32]{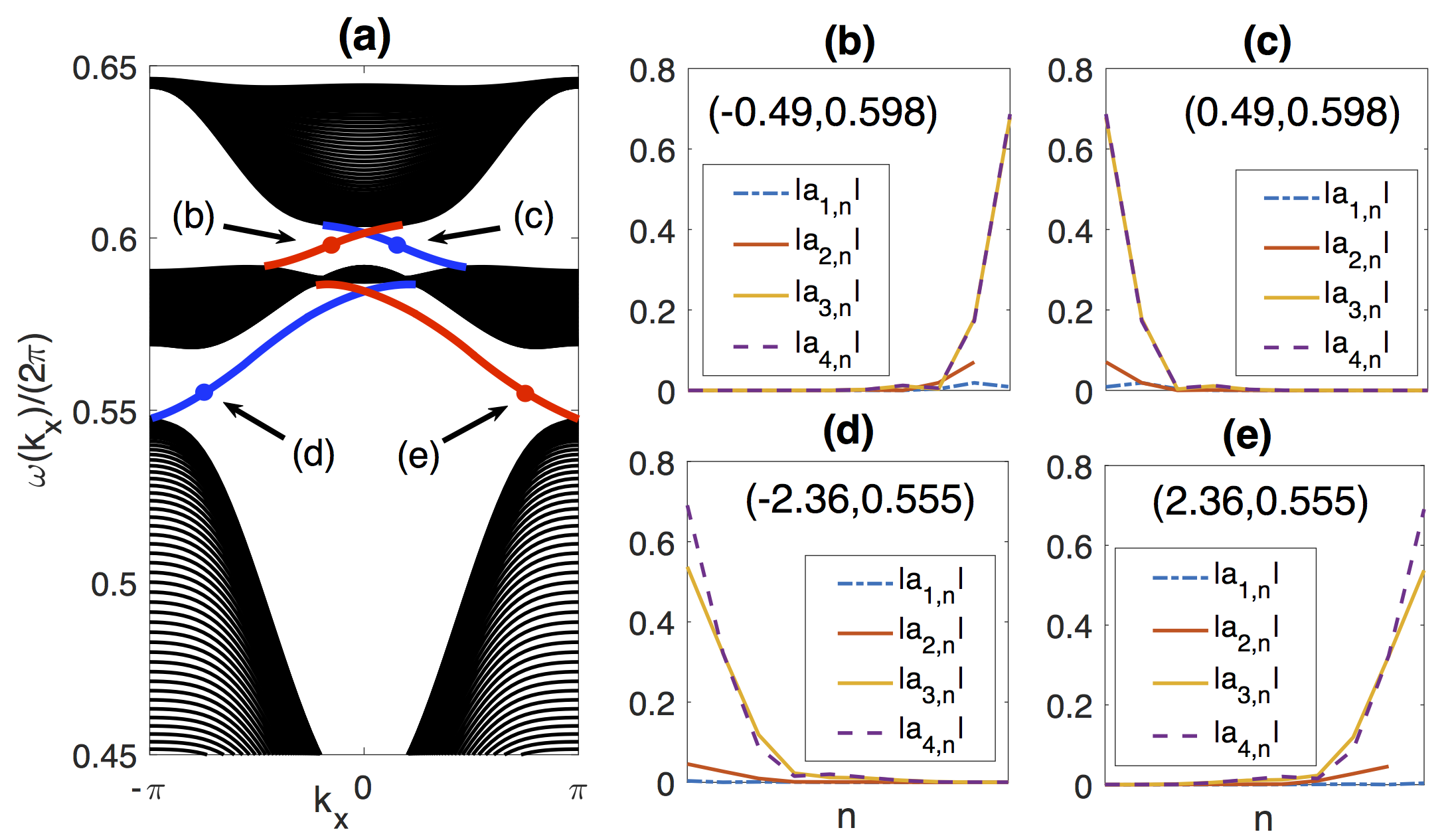}
\caption{(color online) (a) A closer view of the band gaps in Fig.~\ref{edge_bands_plot}(b). A red (blue) curve denotes a family of edge modes localized along the top (bottom) edge. The {magnitude of the} eigenmodes at the point $(k_x, \omega / (2\pi))$ are shown in panels (b-e).}
\label{edge_modes_position}
\end{figure}

Moving up to the band gap between the third and fourth bands the orientation {an envelope will move is} counter-clockwise. We point that this is in the {\it opposite} direction from what was found in the lower gap and will {consequently have an {\it opposite}} chirality sign \cite{AC2}. Overall the width of this zone gap is rather small, but we are still able to identify frequencies that support edges states totally in the gap region. Representative edge states are shown in Fig.~\ref{edge_modes_position}(b) and \ref{edge_modes_position}(c) localized along the top and bottom edges, respectively. Again, near the edge the half-integer eigenmode is found to be small in comparison to the {$p=3,4$}  modes{, {meanwhile} the $p=1$ mode is even smaller.}

There is a direct connection between the bulk Chern numbers and the chirality and number of edge modes. Define the clockwise direction, when viewed from above, as the positive direction. 
Let us denote the band gap between the first and second bulk bands as the $p = 1$ (first) gap, and from there the gap index increases with $\omega$ i.e. $p = 2,3, \dots $ (second, third, ... {gaps}). The bulk-edge correspondence gives the relationship
\begin{equation}
\label{chern_edge_relation}
 \mathcal{I}_{p} = \sum_{q = 1}^{p}  C_q  ~ , 
\end{equation}
where $\mathcal{I}_p$ is the gap Chern number {i.e. the number of gapless modes {(taking into {account the} orientation)}} of the $p^{\rm th}$ band gap and $C_q$ is the Chern number corresponding to the $q^{\rm th}$ {bulk} band. This quantity gives the sum of the (signed) topologically protected modes within that gap. We point out that $\mathcal{I}_p$ is equivalent (up to multiplicative constant) to the Hall conductance considered in the quantum Hall effect \cite{TKNN}. This sum can also be expressed as 
$ \mathcal{I}_p \equiv \mathcal{P}_p - \mathcal{N}_p , $
where $\mathcal{P}_p$ ($\mathcal{N}_p$) is the number of positive (negative) chiral modes in the $p^{\rm th}$ gap. We note that from the gap Chern number alone we are unable to uniquely determine the number of gap edge modes, namely $\mathcal{P}_p$ and $\mathcal{N}_p$.

\begin{table}
\centering
  \begin{tabular}{ | c | c | c | c | c | }
    \hline
   $p$  &   $C_p$ & $\mathcal{I}_p$ &$ \mathcal{N}_p$ & $\mathcal{P}_p$  \\ \hline
    1 & 0 & 0 & 0 & 0 \\ \hline
    2 & 1 & 1 & 0 & 1  \\ \hline
    3 & -2 & -1  & 1 &0 \\ \hline
  \end{tabular}
\caption{ \label{edge_index} Bulk-edge correspondence between bulk and gap Chern numbers. }
\end{table}

The topological values for this system are organized in Table~\ref{edge_index}. Using the Chern numbers shown in Fig.~\ref{edge_bands_plot}(b) we calculate the gap Chern number (\ref{chern_edge_relation}) for the first {three} band gaps. Through examination of the edge bands in Fig.~\ref{edge_modes_position}(a) we are able to verify that the second band gap contains one positive mode and the third gap has one negative chiral mode. An equivalent way of establishing the bulk-edge correspondence is through {the difference} $C_p = \mathcal{I}_p - \mathcal{I}_{p-1}$. If we take $\mathcal{I}_0 \equiv 0$, then this definition is also consistent with the edge band diagram.

\section{Time Evolution}
\label{time_evolve_sec}

In this 
 {section} we consider {dynamics via} time evolution of the tight-binding model. We consider two distinct initial boundary value problems. First, we integrate the system for a wide envelope initial condition. The envelope is observed to move unidirectionally with group velocity and {propagate without backscatter} through lattice defects.  Later we {look at the dynamics due to introducing a delta function source modeling} an input antenna{. We find that the source also} excites unidirectional flow localized along the edge. 

The system of interest is that of the coupled mode {system} in (\ref{tb_eqn_time}). For simplicity the domain is taken to be rectangular. The boundary conditions will consist of integer lattice sites, similar to that shown in Fig.~\ref{draw_TB_BC}(a){, in both directions}. Specifically we impose the following boundary conditions:
\begin{align}
\label{zero_BCs_2d}
 &p = 2: ~~ a_{p,mn}  = 0  \\ \nonumber {\rm for }~~  n \le -N &,~ n > N, ~\text{and}~ m \le -M ,~ m > M ~ , \\ \nonumber
 &p = 1,3,4: ~~ a_{p,mn}  = 0 \\ \nonumber  ~~~~ {\rm for }~~ & |n| > N ~ , ~\text{and}~ |m| > M ~ ,
\end{align}
where {$M,N$ are large integers. 
 {For the computational results shown in Figs.~\ref{elec_field_env}, \ref{modal_field_env}, \ref{source_no_defect_evolve}, and \ref{defect_evolution} we take $N = 25$ and $M = 125$ for the envelope or $M = 75$ for the source.} The system of ODEs is integrated using a fourth order Runge-Kutta method. 
 
 To initialize the envelope function we take 
 \begin{equation}
\label{envelope_IC}
a_{p,mn}(0) = \text{sech}\left[\nu (m - m_0)  \right] a_{p,n}(k_x,\omega) e^{i m k_x} ~,
\end{equation}
{for $p=1,2,3,4$ for the} hyperbolic secant centered at $m = m_0$.
The wavenumber $k_x$ and numerically computed eigenmode $a_{p,n}(k_x,\omega)$ correspond to an eigenvalue that resides in the {gap} 
located between the second and third bulk bands in {Fig.~\ref{edge_modes_position}(a)}. {In particular, we take the frequency that corresponds to the edge mode shown in Fig.~\ref{edge_modes_position}(e).}
 The width parameter $\nu$ is taken to be small so that only gap modes are excited.

In order to probe the robustness of {this mode}, we propagate {it} into {a defect} and monitor the scattering. The defect we wish to model is a region adjacent to the wall through which the electric field can not propagate, and so it must go around or scatter. {This boundary condition is effectively a large potential barrier wall through which the field can not penetrate.} To approximate this scenario we set the field coefficients to be zero at {the lattice sites in this region.} 

\begin{figure} [h]
\centering
\includegraphics[scale=0.4]{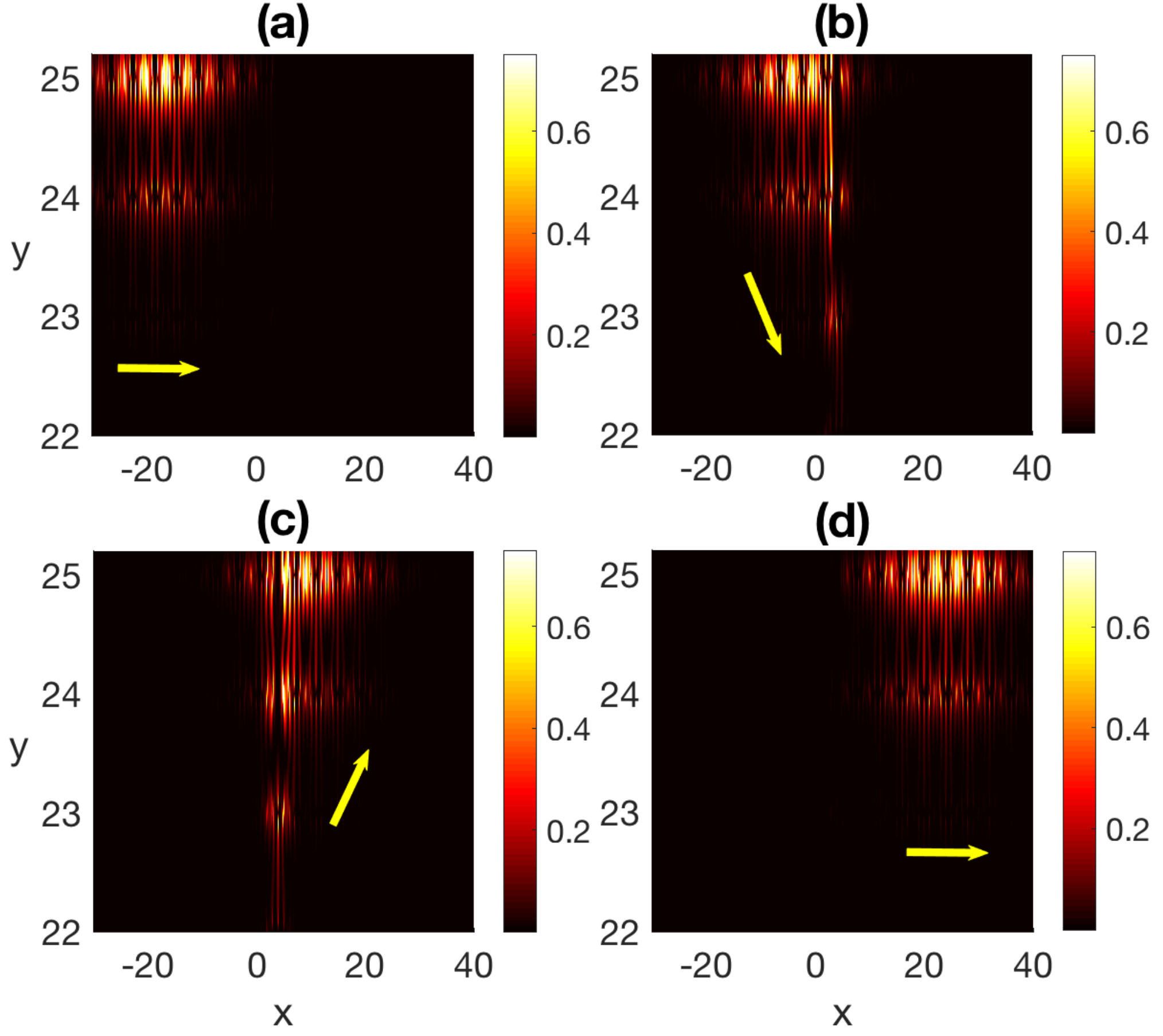}
\caption{Envelope evolution {impacting a defect} for initial condition (\ref{envelope_IC}) with $\nu = 0.125, m_0 = -25$ and {$(k_x , \omega / (2 \pi)) = (2.36,0.555)$}. {Shown is the} intensity  of electric field at (a) $t = 100$, (b) $t =300 $, (c) $t = 480$, and (d) {$t = 700$}.   A lattice defect is imposed at lattice integer sites {$p = 1,3,4, m = 4, 24 \le n \le 25$ and at half-integer sites $p = 2, 4 \le m \le 5, 24 \le n \le 25$}.}
\label{elec_field_env}
\end{figure}

\begin{figure} [h]
\centering
\includegraphics[scale=0.4]{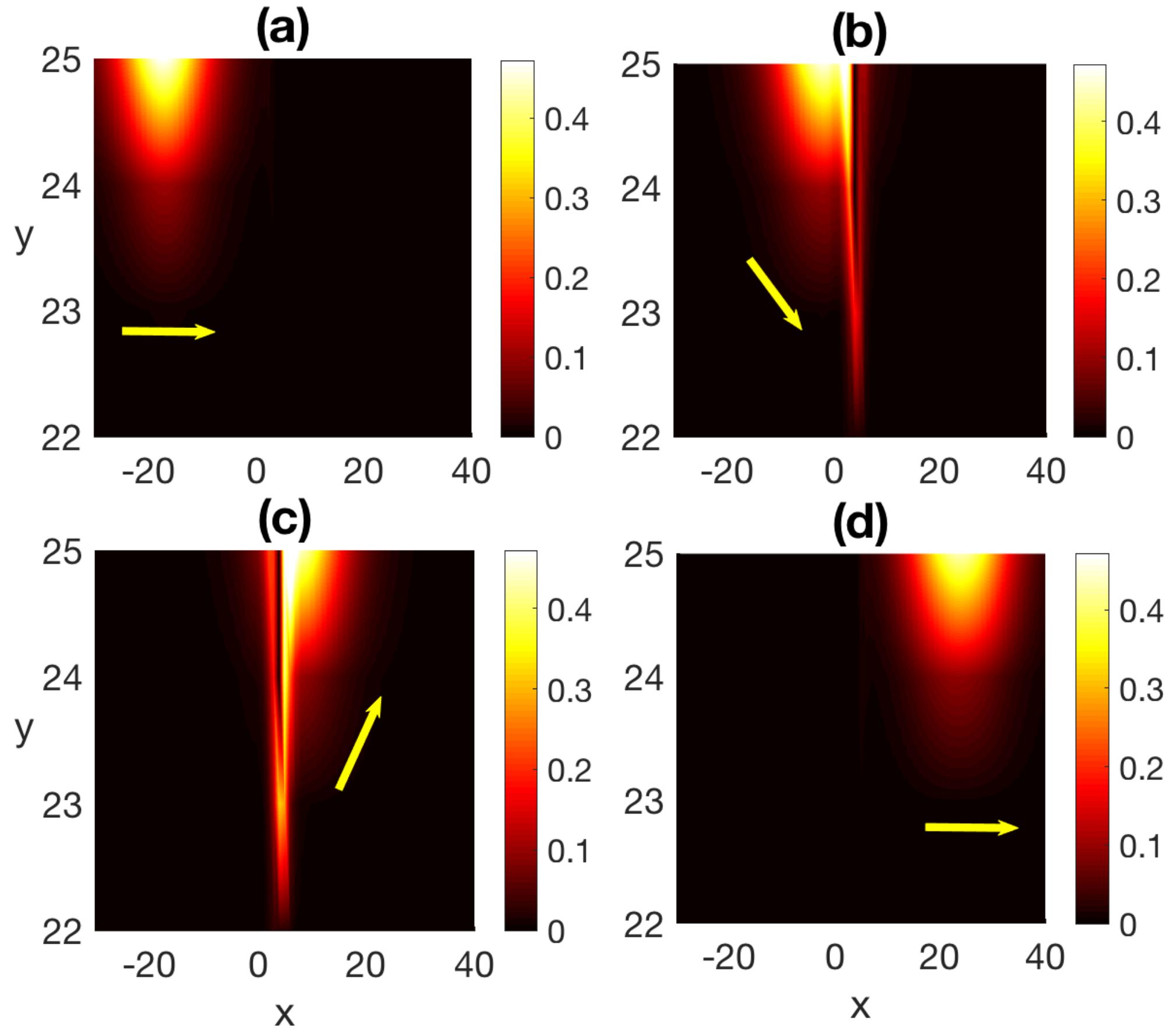}
\caption{Envelope evolution  {impacting a defect for}  initial condition (\ref{envelope_IC}) with $\nu = 0.125, m_0 = -25$ and {$(k_x , \omega / (2 \pi)) = (2.36,0.555)$}. Intensity  of $a_{4,mn}$ mode  at (a) $t = 100$, (b) $t = 300$, (c) $t = 480$, and (d) {$t = 700$}. A lattice defect is imposed at lattice integer sites {$p = 1,3,4, m = 4, 24 \le n \le 25$ and at half-integer sites $p = 2, 4 \le m \le 5, 24 \le n \le 25$}.}
\label{modal_field_env}
\end{figure}

The evolution of the envelope into an edge defect is presented in Figs.~\ref{elec_field_env} and \ref{modal_field_env} which show the  {the {intensity of the} electric field and  {intensity of the} modal envelope, respectively. {The electric field {in Fig.~\ref{elec_field_env}} 
corresponds to the (real-valued) field in (\ref{maxwell}). To find the electric field of the TM equation we use the expansion defined in (\ref{e_field_time_dep}).}  {Shown} 
in Fig.~\ref{modal_field_env} is  the {$ a_{4,mn}$} mode since it is {one of the dominant modes} 
[see Fig.~\ref{edge_modes_position}(e)] and it helps visualize the location of the envelope.} Before encountering the defect the envelope is found to propagate with group velocity {$- \omega'(k_x) \approx 0.0781$} of the edge bands in Fig.~\ref{edge_modes_position}(a). Upon collision with the defect the envelope does {\it not} backscatter. Instead {it} maneuvers around the defect and exits with almost no loss of intensity and the same form it entered with. From this simulation we see that the energy of this system moves (with group velocity) in only one direction. More importantly, without topological protection an envelope mode like this would not possess the robust unidirectionality seen in Figs.~\ref{elec_field_env}-\ref{modal_field_env}.

Next we consider an initially zero field that is excited by a source. 
To incorporate the source term we modify Eq.~(\ref{tb_eqn_time}) to 
\begin{equation}
\label{discrete_eqns_w_source}
\frac{d^2 a_{\ell,mn}}{d t^2} +  \sum_{p = 1}^4     \sum_{\mu,\nu \in {\bf d}_s}    \left[ \mathbb{L}^{\ell p}_{\mu\nu}  + \mathbb{M}^{\ell p}_{\mu\nu} \right] 
 a_{p,m+\mu,n+\nu}  = \frac{d J_{\ell,mn}}{dt}  , 
\end{equation}
 where $ \ell = 1,2,3,4 $ and $J_{\ell, mn}(t) = C_{\ell} \delta_{mq} \delta_{nr} \tanh( bt)e^{i \omega t} $ is {a} 
 time-harmonic source at large times. The point $(q,r)$ indicates the location of the source {(near the top edge)}, while $C_{\ell}$ allows us to control which modes we excite, and the hyperbolic tangent function is included to {gradually} ramp the source up from zero. For the simulation below we look to place an antenna in the region where there is no lattice rods i.e. at a half-integer lattice site. For this reason we take $C_\ell = \delta_{\ell 2} .$ The source is placed {at $(q,r) = (0,25)$} (close to the edge) so that it will excite an edge mode. {In terms of continuous variables,} 
 the spatial location of the source corresponds to  the point $(x,y) = (-0.5, 24.5)$. As an initial condition we take  {$a_{\ell ,mn}(t=0) = 0$}.

\begin{figure} [h]
\centering
\includegraphics[scale=0.38]{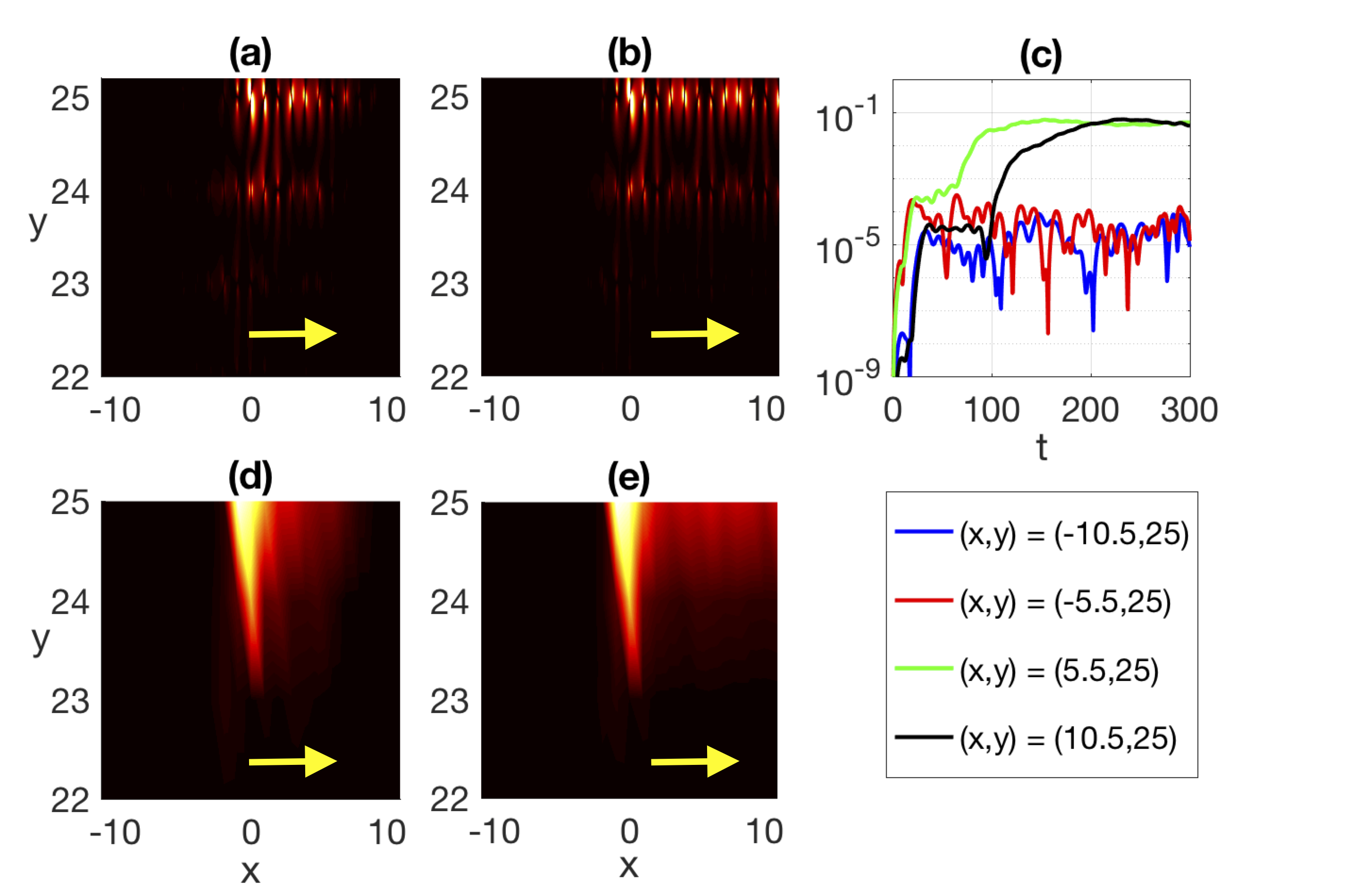}
\caption{(color online) {Evolution of the electric field at frequency $(k_x , \omega/(2\pi)) = (2.36,0.555)$. Snapshots of the intensity at time (a) $t = 100$ and (b) $t = 300$. (c) Time evolution of the electric field {intensity} at points to the left and right of the source. Intensity  of $a_{4,mn}$ mode  at (d) $t = 100$ and (e) $t = 300$.}}
\label{source_no_defect_evolve}
\end{figure}

The evolution of the defect-free problem is summarized in Fig.~\ref{source_no_defect_evolve}. Starting from zero, the source (at a band gap frequency) slowly ramps up to full strength and begins to excite an edge mode. Instead of propagating in all directions the edge state only flows from left to right, unidirectionally. The edge mode slowly begins to fill the wall with a wave front that travels {approximately} with the group velocity. After this early transient period, the system appears to transition into a steady-state {with the electric field concentrated near the {top most} lattice rods.} {To highlight the unidirectional flow of the wave} we measure the field intensity at {points to the left and right of the source}. As shown in {Fig.~\ref{source_no_defect_evolve}(c)}, the difference between the intensity to the right of the source and left of the source is {three to four} orders of magnitude. {Also included in Figs.~\ref{source_no_defect_evolve}(d) and \ref{source_no_defect_evolve}(e) {is} the modal {coefficient $a_{4,mn}(t)$}  {used to compute} 
Figs.~\ref{source_no_defect_evolve}(a) and \ref{source_no_defect_evolve}(b), respectively.}

Next we examine the evolution of a source induced edge mode when it encounters a lattice defect. Using the same initial condition as before, the source slowly excites {the} unidirectional edge state {shown in Fig.~\ref{defect_evolution}}. When the mode encounters the lattice defect it does not backscatter at all. Instead it flows around the defect and eventually rejoins the original wall. {The flow of the field can be seen in Fig.~\ref{defect_evolution}(c) where leftward flow is negligibly small in comparison to the right direction.} The robust unidirectional motion seen here {is similar to}
that seen in {Figs.~\ref{elec_field_env}-\ref{modal_field_env}}
{; {and} is} attributed to {topology; i.e.} the presence of a nonzero gap Chern number. 

\begin{figure} [h]
\centering
\includegraphics[scale=0.5]{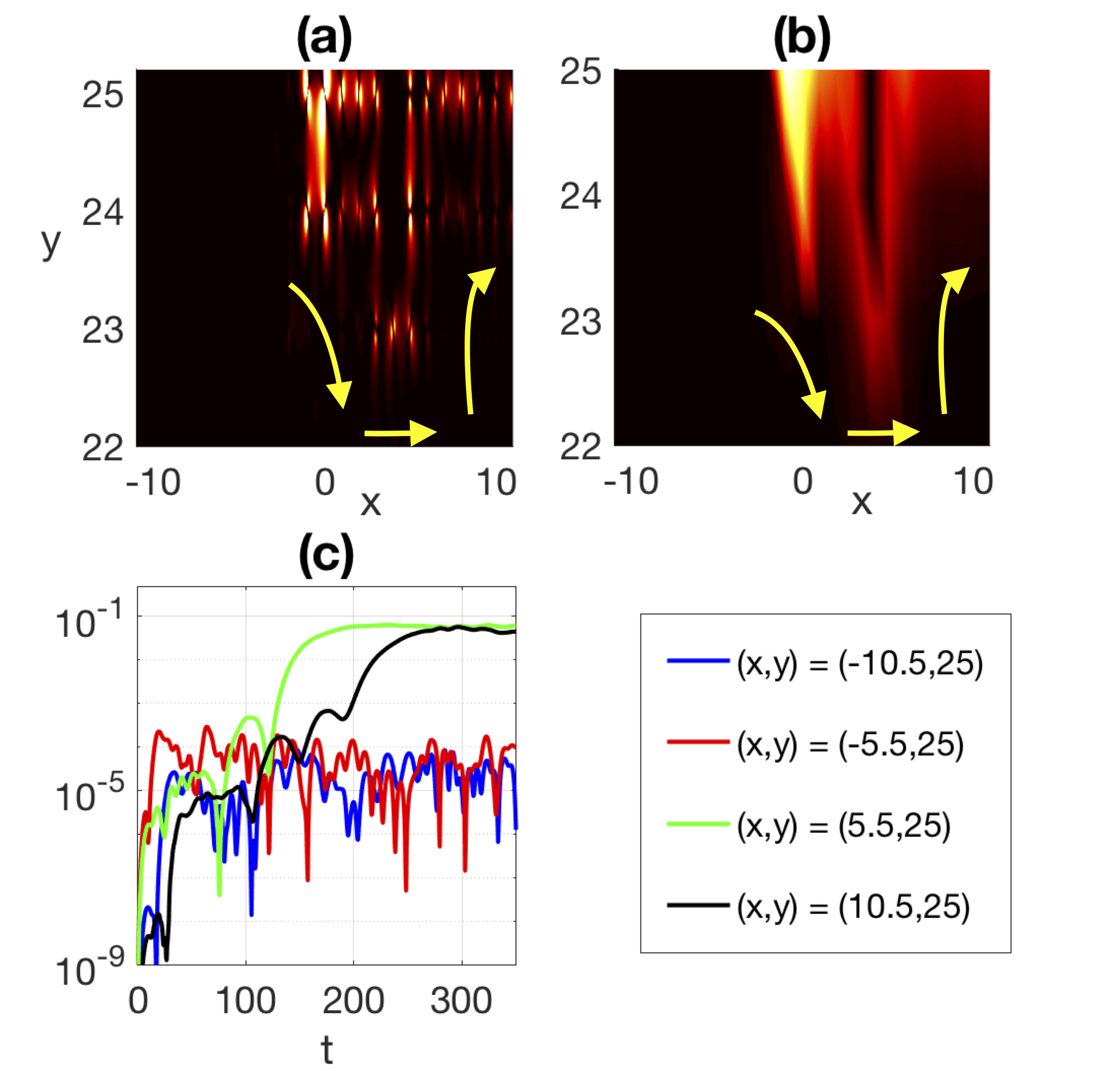}
\caption{(color online) {Evolution of the electric field at frequency $(k_x , \omega/(2\pi)) = (2.36,0.555)$ in the presence of a defect. A lattice defect is imposed at lattice integer sites $p = 1,3,4, m = 4, 24 \le n \le 25$ and at half-integer sites $p = 2, 4 \le m \le 5, 24 \le n \le 25$. (a) Snapshot of the {electric field} intensity at time $t = 350$. (b) Snapshot of the {modal} intensity of $a_{4,mn}$ at $t = 350$. (c) Time evolution of the electric field {intensity} at points to the left and right of the source.}}
\label{defect_evolution}
\end{figure}

\section{Conclusions}
\label{conclude}

{This paper presents the first complete tight-binding  model for the TM Maxwell equation with an external magnetic field that can be used to {effectively} describe the underlying wave dynamics and associated  topological effects. {This discrete model accurately describes the {results in \cite{wang1}}.} The method relies on expansion in terms of suitable Wannier functions. To find these Wannier functions the algorithm in \cite{marzari} was useful.}
Applying the external magnetic field introduces an anisotropic gyrotropic response in the permeability which breaks time-reversal symmetry{. These magnetic effects induce nontrivial topology which in turn} prevents a direct expansion in terms of Wannier modes. {Instead a perturbative approach was used to obtain} a set of Wannier modes  from {the} closely related (time-symmetric) problem {in order to} model the {{full} 
problem which does not have time-reversal {symmetry}}.  

With these Wannier modes a set of tight-binding equations were derived that yield {accurate approximations of the} actual spectral bands. 
 {Importantly,} this discrete system {produces 
 band approximations} with nontrivial Chern number{s and topologically protected edge modes. The Chern numbers/edge modes agree with the bulk-edge correspondence.} 
 {The} edge modes whose frequencies lie in the band gap were found to propagate unidirectionally{, without backscatter} around lattice defects.

 {This} tight-binding model {provides} qualitative agreement with full numerical simulations {at} a fraction of the cost. Solution{s} of the discrete model can be performed on a small laptop in a relatively short amount of time.  This approach serves as an alternative to direct numerics and {paves the way for developing coupled mode models in other problems with nontrivial Chern numbers.} 

\section{Acknowledgement}
This work was partially supported by AFOSR under grant No. FA9550-19-1-0084.

\appendix

\section{The Marzari-Vanderbilt (MV) Algorithm}
\label{MV_alg_sec}

This appendix reviews some of the relevant ideas of the MV algorithm \cite{marzari}. The algorithm is used to compute well-localized Wannier modes that are then used to compute the discrete 
 {coefficients} in Sec.~\ref{TbA_sec}. The algorithm {is needed because} 
the numerically computed Bloch eigenfunctions from (\ref{Eig_Fourier_Form}) {are not sufficiently smooth}
 functions of ${\bf k}$, which {results in Wannier functions (\ref{wannier_define}) with slow decay in ${\bf r}$.} 

First we introduce some preliminary definitions. In \cite{blount} it is shown that the first moment (center of mass) for the $p^{\rm th}$ Wannier mode located near the unit cell (UC): $[-1/2 , 1/2] \times [-1/2 , 1/2]$ is {given  by}
\begin{align}
\label{first_moment}
\overline{{\bf r}}_p \equiv & \langle w_{p,00} | {\bf r} w_{p,00}  \rangle_{\mathbb{R}^2, \epsilon \tilde{\mu}} \\ \nonumber
= & \frac{1}{ 4 \pi^2} \iint_{\rm BZ} \langle u_p({\bf r}, {\bf k}) | \nabla_{\bf k}  u_p({\bf r}, {\bf k}) \rangle_{\rm{UC}, \epsilon \tilde{\mu}} ~d{\bf k} ~ ,
\end{align}
where the above {norm} 
is defined as 
\begin{equation}
\label{norm_define}
||f ||^2_{\rm{UC}, \epsilon \tilde{\mu}} = \iint_{\rm UC} |f({\bf r})|^2 \epsilon({\bf r}) \tilde{\mu}({\bf r}) d{\bf r} ~ ,
\end{equation}
and the second moment of the $p^{\rm th}$-band Wannier mode is 
\begin{align}
\label{second_moment}
\langle |{\bf r}|^2 \rangle_p \equiv & \langle w_{p,00} | |{\bf r}|^2 w_{p,00}  \rangle_{\mathbb{R}^2, \epsilon \tilde{\mu}} \\ \nonumber
=&  \frac{1}{ 4 \pi^2} \iint_{\rm BZ}   ||\nabla_{\bf k}  u_p({\bf r}, {\bf k}) ||^2_{\rm{UC}, \epsilon \tilde{\mu}}~ d{\bf k} ~ . 
\end{align}
Since all other Wannier modes $w_{p,mn}$ are merely translations of the Wannier mode located near the UC, it suffices to focus on optimizing the localization properties of the $w_{p,00}({\bf r})$ modes.
Next introduce the 
spread (variance) functional 
\begin{equation}
\label{spread_func}
\Omega = \sum_{p=2}^4 \Omega_p ~ , ~~~~~~~ \Omega_p =    \langle |{\bf r}|^2 \rangle_p - |\overline{\bf r}_p|^2  ~ .
\end{equation}
The second, third, and fourth bands in Fig.~\ref{spectral_curves} {form} an {isolated} subset of {\it entangled} bands. This means that somewhere throughout the Brillouin zone there are band degeneracies, or touching points, where $\omega_{p} ({\bf k}) = \omega_{p+1} ({\bf k})$.  As such, we must solve for all entangled bands simultaneously, as a coupled system. The goal of the MV algorithm is to minimize (\ref{spread_func}) by introducing a unitary transformation that smoothes the Bloch modes in ${\bf k}$-space.
 
{Consider the linear combination of Bloch functions}  
\begin{align}
\label{quasi_bloch}
\psi_q({\bf r}, {\bf k}) = & \sum_{p = 2}^4 [U_{pq}({\bf k})] E_p({\bf r}, {\bf k}) \\ \nonumber
=&  \sum_{p = 2}^4 [U_{pq}({\bf k})] e^{i {\bf k} \cdot {\bf r}}  u_p({\bf r}, {\bf k}) ~ ,
\end{align}
where $U({\bf k})$ is a {$3 \times 3$}  unitary  matrix whose elements are given by $[U_{pq}({\bf k})]$. {Each of the Bloch functions used in (\ref{quasi_bloch}) are normalized to one with respect to the norm given in (\ref{norm_define}).} By taking a unitary transformation of the original Bloch modes all the orthogonality properties  are inherited by their {linear combination.} The corresponding  Wannier modes are defined in terms of {(\ref{quasi_bloch})} and given by
\begin{equation}
\label{wannier_define2}
w_{p,mn}({\bf r}) = \frac{1}{4 \pi^2} \iint_{\rm BZ} e^{- i {\bf k} \cdot {\bf R}_{mn}} \psi_p({\bf r}, {\bf k}) ~  d{\bf k} ~ .
\end{equation}
This is the formula { {used} to calculate} the Wannier modes {shown} in Fig.~\ref{wannier_modes_plot}.

 Next we discuss the MV algorithm approach to numerically {compute} Wannier modes that minimize spread.
Start by discretizing the Brillouin zone $[-\pi , \pi] \times [-\pi ,  \pi]$ by the mesh ${\bf k}_{m'n'} = \left(- \pi + \Delta k \left(m' +  \frac{1}{2} \right), - \pi + \Delta k \left(n' +  \frac{1}{2} \right) \right)$ for $m',n' = 0,1,2, \dots, P-1$, where $\Delta k = 2 \pi/ P.$ 
Integrals are replaced by trapezoidal quadratures 
$$ \frac{1}{4 \pi^2} \iint_{\rm BZ} d{\bf k} \rightarrow \frac{1}{4 \pi^2} \sum_{\bf k} (\Delta k)^2 =  \frac{1}{P^2} \sum_{\bf k}  ~ ,$$
where summation is over all points in the mesh. For functions with periodic boundary conditions in ${\bf k}$ this quadrature is exponentially accurate. Next the derivatives in (\ref{first_moment}) and (\ref{second_moment}) are approximated by second-order accurate and centered finite-differences stencils.
The gradient approximation is 
\begin{equation*}
\nabla_{\bf k} f({\bf k}) \approx \sum_{{\bf b} \in \mathcal{S}} w_b {\bf b} \left[ f({\bf k} + {\bf b}) - f({\bf k}) \right] ~ ,
\end{equation*}
over  the set 
\begin{equation}
\mathcal{S} = \left\{ \begin{pmatrix} \Delta k \\ 0 \end{pmatrix} , \begin{pmatrix} -\Delta k \\ 0 \end{pmatrix} , \begin{pmatrix} 0 \\ \Delta k \end{pmatrix} , \begin{pmatrix} 0 \\ - \Delta k \end{pmatrix} \right\} ~ ,
\end{equation}
 with $w_b = 1/( 2 \Delta k^2)$ for each vector. {The following  discrete approximations to the first and second moments given in (\ref{first_moment}) and (\ref{second_moment}), respectively,  are {derived} in \cite{marzari} }
\begin{equation}
\label{COM_discrete2}
\overline{{\bf r}}_p \approx - \frac{1}{P^2} \sum_{{\bf k}, {\bf b}} w_b {\bf b} {\rm Im} \log [M_{pp}{({\bf k}, {\bf b})}] ~ ,
\end{equation}
and
\begin{align}
\label{SecMom_discrete2}
\langle |{\bf r}|^2 \rangle_p \approx  \frac{1}{P^2} \sum_{{\bf k}, {\bf b}} w_b  \bigg\{ & - 2 {\rm Re} \log [M_{pp} {({\bf k}, {\bf b})}]  \\ \nonumber +& \left({\rm Im} \log  \left[ M_{pp}{({\bf k}, {\bf b})} \right] \right)^2 \bigg\} ~ ,
\end{align}
 {where}
\begin{equation}
\label{define_m_matrix}
[M_{pq}({\bf k}, {\bf b})] = \langle u_p({\bf r}, {\bf k}) |  u_q({\bf r}, {\bf k}+ {\bf b}) \rangle_{\rm{UC}, \epsilon \tilde{\mu}} ~ 
\end{equation}
 {are the elements of the {matrix} $M({\bf k}, {\bf b})$.}
We point out that this inner-product is performed only once at the beginning of the algorithm below. Afterward updates are applied directly to the matrix $M({\bf k}, {\bf b})$. Since this matrix does not depend on space, the algorithm iterates quite fast. We also note that $M({\bf k}, {\bf b})$ is periodic in ${\bf k}$.

In order to minimize spread functional (\ref{spread_func}) a gradient descent optimization algorithm is implemented and used  to update the unitary matrix in (\ref{quasi_bloch}). The descent gradient in \cite{marzari} is  found to be 
\begin{equation}
\label{gradient_define}
\Delta W({\bf k}) = \frac{\alpha}{w} \sum_{\bf b} w_b \left[ \mathcal{A}[\tilde{R}{({\bf k},{\bf b})}] -\mathcal{S}[T{({\bf k},{\bf b})}]\right]   ~ ,
\end{equation}
where $w = \sum_{\bf b} w_b$ and
\begin{eqnarray*}
&  [T_{pq}{({\bf k}, {\bf b})}]  = [\widetilde{R}_{pq}{({\bf k}, {\bf b})}] [Q_q{({\bf k}, {\bf b})}] \; ,  & \\\\
& [\widetilde{R}_{pq}{({\bf k}, {\bf b})}]  = \frac{[M_{pq}{({\bf k}, {\bf b})}]}{ [M_{qq}{({\bf k}, {\bf b})}] } \; , & \\ 
&  [Q_q{({\bf k}, {\bf b})}]  = {\rm Im} \ln [M_{qq}{({\bf k}, {\bf b})}] + {\bf b} \cdot \overline{\bf r}_q \; .  &
\end{eqnarray*}
The operators $\mathcal{A}[B]$ and $\mathcal{S}[B]$ are defined by
\begin{equation*}
\mathcal{A}[B] = \frac{B - B^\dag}{2} \; , ~~~~~~ \mathcal{S}[B] = \frac{B + B^\dag}{2i} \; ,
\end{equation*}
where $\dag$ denotes the complex conjugate transpose. By construction, the matrix $\Delta W$ is anti-unitary i.e. $\Delta W^\dag = - \Delta W,$ and it preserves time-reversal symmetry, that is $\Delta W(-{\bf k}) = \Delta W({\bf k})^*$. So, as a result, if an initial guess of the algorithm has time-reversal symmetry, then so will the output.


The value of $\alpha $ used in (\ref{gradient_define}) is taken to be small and positive, typically 0.1 in our simulations. The unitary matrix in (\ref{quasi_bloch}) is updated by 
\begin{equation}
\label{Unitary_iterate}
U^{(c+1)}({\bf k}) = U^{(c)}({\bf k}) \exp\left( \Delta W^{(c)}({\bf k}) \right) \; ,
\end{equation}
where $c$ denotes the iteration count, starting with the initial guess at $c = 0$.
By construction, this update is also a unitary matrix since $\Delta W$ is anti-unitary.
The matrix in (\ref{define_m_matrix}) is then updated by 
\begin{equation}
\label{overlap_iterate}
M^{(c+1)}({\bf k}, {\bf b}) = (U^{(c+1)}({\bf k}))^\dag M^{(0)}({\bf k}, {\bf b}) U^{(c+1)}({\bf k} + {\bf b}) \; .
\end{equation}
To obtain the Wannier modes shown in Fig.~\ref{wannier_modes_plot} we used the algorithm above with the initial guess described in the next section. 
{A summary of the iteration method is given below.
\begin{enumerate}
\item Numerically compute the Bloch modes  from Eq.~(\ref{master_eqn_bloch_sub}).
\item Rescale the Bloch mode as shown in Eq.~(\ref{scalar_cancel}).
\item Compute initial overlap matrix $M^{(0)}({\bf k}, {\bf b})$ in Eq.~(\ref{define_m_matrix}) using the rescaled Bloch modes.
\item Initialize unitary matrix $U^{(0)}({\bf k}) $ as {an} identity matrix.
\item Begin iteration loop, starting at $c = 0$:
\begin{enumerate}
\item Compute anti-unitary matrix {$\Delta W^{(c)}({\bf k})$} in Eq.~(\ref{gradient_define}).
\item Update the unitary matrix $U^{(c+1)}({\bf k})$ in Eq.~(\ref{Unitary_iterate}).
\item Update overlap matrix $M^{(c+1)}({\bf k}, {\bf b})$ in Eq.~(\ref{overlap_iterate}).
\item Stop iteration {when the} difference in successive spreads $|\Omega^{(c+1)}- \Omega^{(c)} |$ in Eq.~(\ref{spread_func}) is less than {a specified} tolerance.
\end{enumerate}
\end{enumerate}

}
 
We remark that the first spectral band in the time-reversal {\it broken} problem [see Fig.~\ref{spectral_curves}(c)] is isolated and has trivial Chern number. 
As a result, after the renormalization procedure described below, the Bloch mode requires no minimization through the MV algorithm to achieve a well-localized Wannier state. Hence the first spectral mode is computed by itself. Only the Wannier modes corresponding to the second, third, and fourth spectral bands are optimized using the MV algorithm. 

\subsection{{An Initial {Bloch function}  through {rescaling}}} 
\label{initial_guess_sec}

The MV algorithm above requires a reasonable initial guess in order to converge to a well-localized mode.
The numerically computed {normalized} Bloch mode is non-unique and has the form {$\mathcal{E}({\bf r}, {\bf k}) = C({\bf k}) E({\bf r}, {\bf k})$} due to the linearity of (\ref{master_eqn_non_dim}). {The function $E({\bf r}, {\bf k})$ is a Bloch mode solution that is assumed to be {sufficiently} 
smooth in {\bf k}.} The pre-factor part is typically {non-smooth in {\bf k}} 
 and can be removed by a simple rescaling of the eigenfunction. If we divide {$\mathcal{E}({\bf r},{\bf k})$} by the same function evaluated at some spatial point ${\bf r}_0$ where  {$\mathcal{E}({\bf r}_0,{\bf k}) \not= 0 $}, then the pre-factor part 
 of the numerical solution is canceled out and we get {
\begin{equation}
\label{scalar_cancel}
\frac{\mathcal{E}({\bf r}, {\bf k}) }{\mathcal{E}({\bf r}_0, {\bf k}) } = \frac{ C({\bf k}) E({\bf r}, {\bf k})}{C({\bf k}) E({\bf r}_0, {\bf k})} = \frac{ E({\bf r}, {\bf k})}{E({\bf r}_0, {\bf k})} \; ,
\end{equation}
} which is also a solution to (\ref{master_eqn_non_dim}). After doing this the Bloch function needs to renormalized so that the {norm of the Bloch mode (\ref{norm_define}) is one.} We find that this typically leads to a smoother initial guess. In our experience it is possible without too much searching to find an ${\bf r}_0$ value that leads to convergence {mentioned} below.

In addition to smoothing the electric field, this rescaling method also preserves the time-reversal symmetry of the $\mathcal{M}({\bf r}) = {\bf 0}$ problem. Notice that {
\begin{equation}
\frac{\mathcal{E}^*({\bf r}, -{\bf k}) }{\mathcal{E}^*({\bf r}_0, -{\bf k}) } = \frac{ E^*({\bf r}, -{\bf k})}{E^*({\bf r}_0, -{\bf k})} = \frac{ E({\bf r}, {\bf k})}{E({\bf r}_0, {\bf k})} = \frac{\mathcal{E}({\bf r}, {\bf k}) }{\mathcal{E}({\bf r}_0, {\bf k}) }  \; .
\end{equation}
 } {We have found it convenient (not necessary) to} divide the first (isolated), second and fourth modes by the Bloch wave at the origin, 
while we divide the third band mode by some nonzero {$\mathcal{E}({\bf r}_0\not= {\bf 0} , {\bf k})$} value. 

\subsection{Convergence Results}
\label{MV_converge_sec}

In this section we apply the MV algorithm given above and {discuss} the convergence results. The Bloch functions are numerically computed using the spectral method described in Sec.~\ref{num_comp_spec_bands}. For these results we have discretized the Brillouin zone with a $24 \times 24$ mesh. To initialize the MV algorithm we use the rescaled initial guess in Sec.~\ref{initial_guess_sec}. The convergence of the method is tracked by calculating the difference in successive iterations of the total spread (\ref{spread_func}). When the difference $|\Omega^{(c)} - \Omega^{(c-1)} |$, where $c$ is the number of iterations, is less than {$ 10^{-10}$} we terminate the algorithm.

The convergence of the MV algorithm is shown in Fig.~\ref{MV_convergence_results} as a function of the number of iterations. For the parameters and initial condition used here, the method takes {14,275} iterations to converge. The variances  for the individual  modes in (\ref{spread_func}) are:  $\Omega_2 = 0.3291$, $\Omega_3 = 0.0330$, and $\Omega_4 = 0.0330$. The  $p =2$ mode is observed to have the largest spread; this mode corresponds to the off-site Wannier mode shown in {Fig.~\ref{wannier_modes_plot}(b)}. The other two Wannier modes have better localization and are centered at lattice sites in {Figs.~\ref{wannier_modes_plot}(c-d)}. Finally, to obtain $ p = 1$ mode we only performed the rescaling in Sec.~\ref{initial_guess_sec} and did not apply the MV algorithm. Doing this produced a mode with variance {$\Omega_1 = 0.0445$}. 

\begin{figure}
\centering
\includegraphics[scale=0.5]{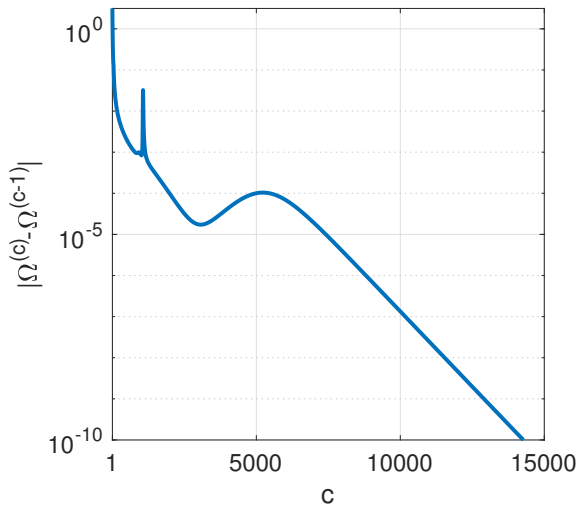}
\caption{Convergence of the spread functional (\ref{spread_func}) for the three band case ($p = 2,3,4$).}
\label{MV_convergence_results}
\end{figure}

\section{Edge Modes with Different Boundary Conditions}
\label{diff_BC_sec}

{This} section is dedicated to showing how altering the boundary conditions {can significantly change structure of the edge band diagram{, while the}  number of edge modes remains the same}. To begin, recall in Fig.~\ref{edge_modes_position}(a) the edge bands were computed for boundaries that began and ended with integer lattice rows [see Fig.~\ref{draw_TB_BC}(a)]. However this is not the only boundary {conditions} we can {introduce}. Another option is to impose a top or bottom row that is all half-integer sites [see Figs.~\ref{draw_TB_BC}(b) and \ref{draw_TB_BC}(c)]. Physically, we interpret this as to where the location wall boundary condition (solid black lines in Fig.~\ref{draw_TB_BC}) {is placed} relative to the integer or half-integer sites. {Recall that the integer sites correspond to the location of ferrite rods.}

\begin{figure} [h]
\centering
\includegraphics[scale=0.55]{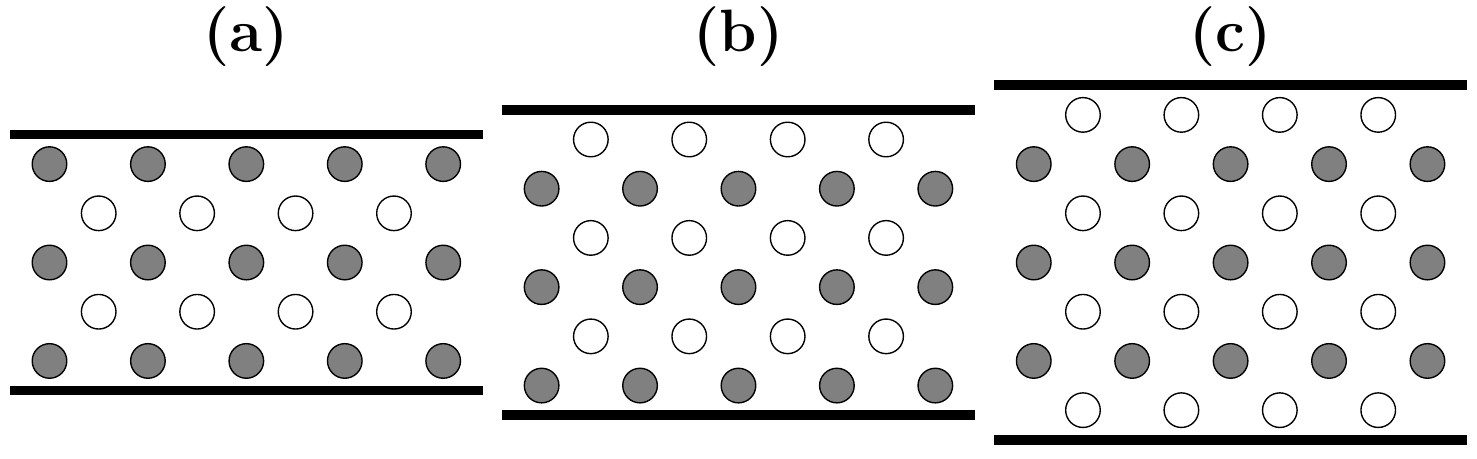}
\caption{Different boundary condition configurations. Gray (white) circles denote integer (half-integer) lattice sites.}
\label{draw_TB_BC}
\end{figure}

The band diagrams corresponding to these edge configurations are shown in Fig.~\ref{edge_bands_compare_BC}. Notice that Fig.~\ref{edge_bands_compare_BC}(a) was considered in Sec.~\ref{discrete_edge_bands_sec}. The band diagrams displayed in Figs.~\ref{edge_bands_compare_BC}(b-c) contain {one (b) or two (c) rows of}
half-integer lattice sites. The most significant difference is the large number of edge eigenvalues that {share frequencies} with bulk modes in the frequency range {$(0.448,0.547)$}. {Two things} 
that {do} 
not change with boundary conditions{: (a)} the presence of a unidirectional state located in the zone gap region of the diagram{, and (b) the bulk bands}.

The curvature and shape  of the dispersion curves corresponding to edge modes have sensitive dependence on the boundary conditions (see also \cite{AC2}). While the edge modes vary with boundary conditions, the bulk bands are found to be impervious to changes at the edges. Moreover the Chern numbers, calculated from the bulk bands, do not depend on the edges and with them the number of gapless edge states. Once a bulk band acquires a nontrivial Chern number, there will be a topological edge mode present  even when defects along the boundary are introduced. 

\begin{figure} [h]
\centering
\includegraphics[scale=0.32]{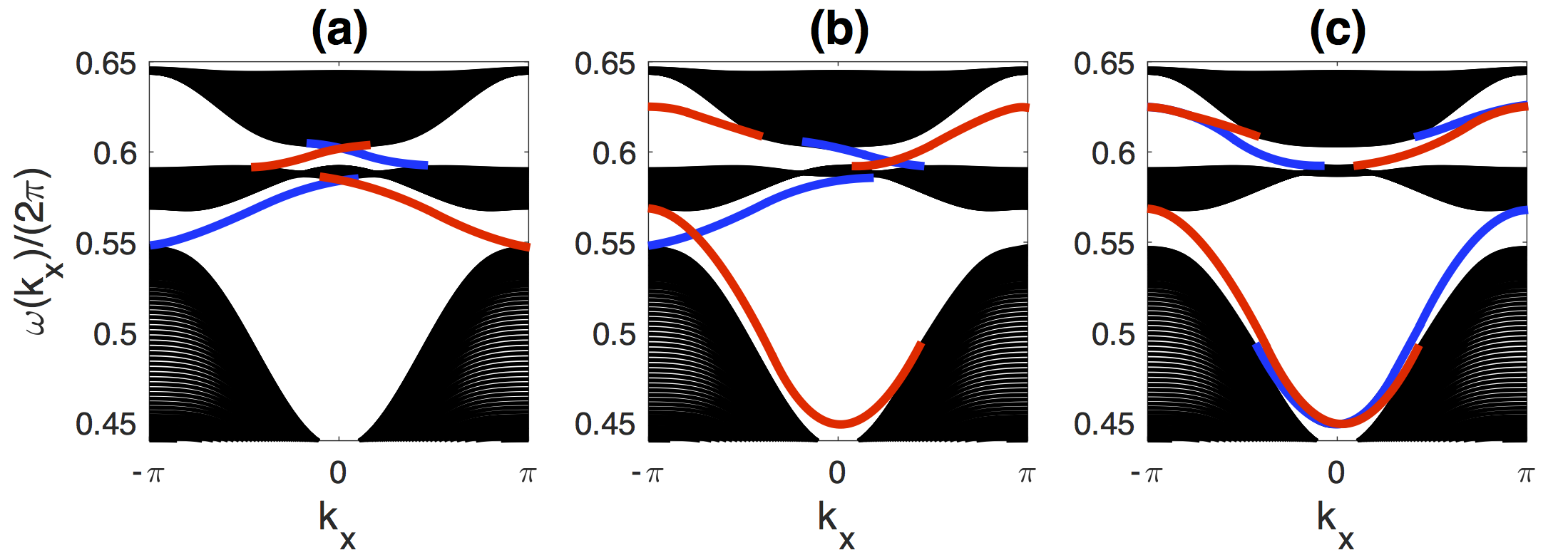}
\caption{(color online) Dispersion bands corresponding to the boundary conditions shown in Fig.~\ref{draw_TB_BC}, respectively. Red (blue) curves correspond to edge modes localized along the top (bottom) edge.}
\label{edge_bands_compare_BC}
\end{figure}

\section{{Improving the Tight-binding Approximation}}
\label{asym_limit_tba}

A natural question to ask is how could the discrete model better approximate the system? {One way to accomplish this is to increase the number of Wannier mode interactions.} In Sec.~\ref{spec_band_sec} we computed the spectral bands and included all interactions with nearby Wannier modes whose center was a euclidean distance of {one} or less away. We can increase the number of interactions in the model for an overall better approximation of the problem, up to a point. On the other hand however, more interactions leads to a higher complexity and will become computationally expensive to solve. In Sec.~\ref{spec_band_sec} we {opted for the fewest number of interactions that gave us a reasonable set of topological bands.} 

\begin{figure} [h]
\centering
\includegraphics[scale=0.29]{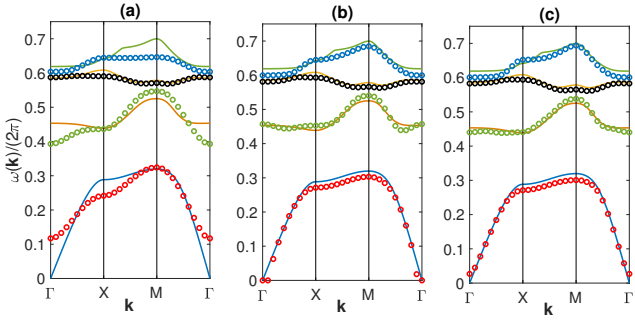}
\caption{The spectral bands  over the irreducible Brillouin zone for (a) ${\bf d}_1$, (b) ${\bf d}_2$, and (c) ${\bf d}_3$. {Note that panel (a) is the same as Fig.~\ref{MO_spec_bands_compare}(b)}}
\label{compare_spec_bands_num_wan_mode}
\end{figure}

A comparison of the spectral bands for different numbers of Wannier mode interactions is shown in Fig.~\ref{compare_spec_bands_num_wan_mode}{, where only the real part is shown.} As the number of interactions increases, qualitatively the discrete approximation is found to improve. This improvement is quantified in Table~\ref{num_wannier_modes_comp} where the relative {band} error and the {inversion} symmetry error are given. Among the $p = 2,3,4$ bands, the {relative} max error is {$11.5 \%, 3.3 \%$ and $2.6 \%$}, respectively, in Figs.~\ref{compare_spec_bands_num_wan_mode}(a), (b), and (c). The table also shows that all three cases have excellent preservation of inversion symmetry. We also point out that all three of these band configurations {have} the {nontrivial} Chern numbers shown in Fig.~\ref{MO_spec_bands_compare}(b). Furthermore, when solved with edge boundary conditions, each set supports gapless edge modes that are topologically protected.

\begin{widetext}
\begin{table}
\centering
  \begin{tabular}{| c |c | c | c | c | c | c | }
    \hline  \multirow{4}{*}{${\bf d}_1$}  & 
   $p$ & 1 & 2 &3 & 4 \\ \hline &
    $\frac{|| \omega_{\rm num}({\bf k})  - \omega_{\rm disc}({\bf k})||_{\max}}{|| \omega_{\rm num}({\bf k}) ||_{\max}}$ & $36.62 \times 10^{-2}$  & $11.50 \times 10^{-2}$ & $2.96 \times 10^{-2}$   &  $7.66 \times 10^{-2}$  \\  \cline{2-6} &  
    $|| \omega_{\rm disc}({\bf k}) - \omega_{\rm disc}(-{\bf k}) ||_{\max}$  & $1.69 \times 10^{-5}$   & $1.09 \times 10^{-4}$  
   & $7.38 \times 10^{-5}$  & $8.50 \times 10^{-5}$  \\ \hline  \hline  \multirow{2}{*}{${\bf d}_2$}  &  
    $\frac{|| \omega_{\rm num}({\bf k})  - \omega_{\rm disc}({\bf k})||_{\max}}{|| \omega_{\rm num}({\bf k}) ||_{\max}}$ &$ 18.02 \times 10^{-2}$  & $ 3.31 \times 10^{-2}$ & $ 2.54 \times 10^{-2}$   &2.85  $ \times 10^{-2}$  \\ \cline{2-6}  &
    $|| \omega_{\rm disc}({\bf k}) - \omega_{\rm disc}(-{\bf k}) ||_{\max}$ & $1.56 \times 10^{-5}$  & $6.12 \times 10^{-5}$ & $2.76 \times 10^{-5}$  & $3.46  \times 10^{-5}$  \\ \hline \hline  \multirow{2}{*}{${\bf d}_3$}  &
    $\frac{|| \omega_{\rm num}({\bf k})  - \omega_{\rm disc}({\bf k})||_{\max}}{|| \omega_{\rm num}({\bf k}) ||_{\max}}$ & $8.42 \times 10^{-2}$  & $2.53 \times 10^{-2}$  & $2.39 \times 10^{-2}$ & $2.62 \times 10^{-2}$   \\ \cline{2-6}  &
    $|| \omega_{\rm disc}({\bf k}) - \omega_{\rm disc}(-{\bf k}) ||_{\max}$ & $6.71 \times 10^{-6}$  & $2.89 \times 10^{-5}$ & $1.17 \times 10^{-5}$  & $1.39 \times 10^{-5}$  \\ \hline
  \end{tabular}
\caption{ \label{num_wannier_modes_comp} Comparison between discrete (disc) and numerical (num) spectral bands in Fig.~\ref{compare_spec_bands_num_wan_mode}. The max norm is defined by $|| f({\bf k})  ||_{\max} \equiv \max_{{\bf k} \in BZ}  |f({\bf k})| $.}
\end{table}
\end{widetext}

\end{document}